\documentclass[times]{elsarticle}

\usepackage{amssymb}
\usepackage{hyperref}
\usepackage{geometry}
\usepackage{psfrag,color,epsfig}
\usepackage{graphicx,graphics}	
\usepackage{multirow}
\usepackage{tabularx}  
\usepackage{url}
\usepackage{xspace}
\usepackage{subcaption}
\usepackage[dvipsnames]{xcolor}
\hypersetup{colorlinks=true,
            linkcolor=blue,
            urlcolor=blue,
            linktoc=all,
            citecolor=blue}
\def\mpci{~{\rm Mpc}^{-1}}
\def\pks{[P(k)]_{\rm SDSS}}
\def\pkr{[P(k)]_{\rm R}}
\def\pkb{[P(k)]_{\rm B}}
\def\bks{[B(k_1,\mu,t)]_{\rm SDSS}}
\def\bkr{[B(k_1,\mu,t)]_{\rm R}}
\def\bkb{[B(k_1,\mu,t)]_{\rm B}}

\journal{New Astronomy}

\begin{document}

\begin{frontmatter}


\title{The size and shape dependence of the SDSS galaxy bispectrum }

\author[first]{Anindita Nandi}
\affiliation[first]{organization={Department of Physics, Visva-Bharati University},
            city={Santiniketan},
            postcode={731235}, 
            country={India}}
\ead{anindita.nandi96@gmail.com}

\author[second]{Sukhdeep Singh Gill}
\affiliation[second]{organization={Department of Physics, Indian Institute of Technology Kharagpur},
            postcode={721302}, 
            country={India}}

\author[third,fourth]{Debanjan Sarkar}
\affiliation[third]{organization={Department of Physics, Ben-Gurion University of the Negev},
            addressline={}, 
            city={Be’er Sheva},
            postcode={84105}, 
            country={ Israel}}

\affiliation[fourth]{organization={Department of Physics and Trottier Space Institute, McGill University},
            addressline={QC H3A 2T8}, 
            country={Canada}}

\author[fifth]{Abinash Kumar Shaw}
\affiliation[fifth]{organization={Astrophysics Research Center of the Open University (ARCO), The Open University of Israel},
            addressline={1 University Road}, 
            city={Ra’anana},
            postcode={4353701}, 
            country={ Israel}}

\author[first]{Biswajit Pandey}
\author[second]{Somnath Bharadwaj}

\begin{abstract}
We have measured the spherically averaged bispectrum of the SDSS main galaxy sample, considering a volume-limited  $[296.75\, \rm Mpc]^3$ data cube with mean galaxy number density   $0.63 \times 10^{-3} \, {\rm Mpc}^{-3}$ and median redshift  $0.102$. Our analysis considers $\sim 1.37 \times 10^{8}$  triangles, for which we have measured the binned bispectrum and analysed its dependence on the size and shape of the triangle. It spans wavenumbers  $k_1=(0.075-0.434)\,{\rm Mpc}^{-1}$ for equilateral triangles, and a smaller range of $k_1$ (the largest side) for triangles of other shapes. For all shapes, we find that the measured bispectrum is well modelled by a power law $A\,\big(k_1/1\mpci\big)^{n}$, where the best-fit values of $A$ and $n$ vary with the shape. The parameter $A$ is the minimum for equilateral triangles and increases as the shape is deformed to linear triangles where the two largest sides are nearly aligned,  
 reaching its maximum value for $\mu = 0.95,\, t = 0.75$.  The values of $n$ are all negative, $|n|$ is minimum $(3.12 \pm 0.35)$ for the shape bin $\mu = 0.65,\, t = 0.75$, and $3.8 \pm 0.28$  for $\mu = 0.65,\, t = 0.85$.
We have also analysed mock galaxy samples constructed from $\Lambda$CDM N-body simulations by applying a simple Eulerian bias prescription where the galaxies reside in regions where the smoothed density field exceeds a threshold. We find that the bispectrum from the mock samples with bias $b_1=1.2$ is in good agreement with the SDSS results. We further divided our galaxy sample into red and blue classes and studied the nature of the bispectrum for each category. The red galaxies exhibit higher bispectrum amplitude $A$ than the blue galaxies for all possible triangle configurations. Red galaxies are old, and their larger bispectra indicate non-linear evolutionary interactions within their environments over time, resulting in their distribution being highly clustered and more biased than younger blue galaxies.
\end{abstract}


\end{frontmatter}




\section{Introduction}
\label{intro}
Observations of the spatial distribution of galaxies  enable us to study  the large-scale structures (LSS) in the Universe.  A variety of works  \citep{sefu09,fergusson2012,Oppizzi2018,shiraishi2019,Feldman2001,Liguori2010},  and  particularly the Planck 2018  results  \citep{planck2020}, all indicate that the LSS originated from very small amplitude (linear) Gaussian primordial density perturbations. 
It is possible to entirely quantify the statistics of these primordial perturbations using the two-point correlation function or its Fourier counterpart the  power spectrum.  The subsequent amplification of these perturbations, through the process of gravitational instability \citep{1980Peebles}, is non-linear and the LSS  ceases to be Gaussian (e.g. \cite{1984Fry,1994PhRvLFry,1994Bharadwaj}).  The three-point correlation function or its Fourier counterpart the bispectrum are the lowest order statistics which quantify  the non-Gaussianity in the LSS (e.g. \cite{1982Fry}).  The bispectrum  $B(k_1,k_2,k_3)$, which refers to a closed triangle of sides   $(k_1,k_2,k_3)$ respectively, depends on both the shape and size of the triangle. 
It is possible to use the triangle-shape dependence of the bispectrum to measure 
 cosmological parameters like the matter density parameter $\Omega_{m0}$, and the bias \citep{1995Hivon,1998MNRASVerde,1997MNRASMatarrese,1999Taruya,1999ApJScoccimarro}.

An initial study \citep{1982Fry} measured the bispectrum of the Lick galaxy catalog, while a subsequent study \citep{1999Frieman}  measured the three-point correlation of the APM survey, both of which are angular sky surveys. \citet{1998Jing} have measured the three-point correlation function of the galaxies in the Las Campanas Redshift Survey, which is a thin nearly two-dimensional survey.  In the first works which have carried out fully three-dimensional analyses, \citet{2001Scoccimarro} and subsequently \citet{Feldman2001} have measured the bispectrum of the galaxies in the IRAS Redshift Catalogs. They have quantified the shape dependence of the bispectrum, and used this to estimate $b_1$ and $b_2$ the linear and quadratic bias parameters respectively. The IRAS PSC$z$ catalog analysed in \citet{Feldman2001} contains $13,180$ galaxies in the range $20 \, h^{-1} {\rm  Mpc} \le 
r \le  500 \, h^{-1} {\rm Mpc}$ distributed in the region of the sky with galactic latitude 
$\mid b \mid \ge 10^{\circ}$, and they have considered triangles  whose sides vary in the range $0.05 \, h \, {\rm Mpc}^{-1}$ to $0.4 \, h \, {\rm Mpc}^{-1}$. \citet{2002Verde} have measured the bispectrum of the 2dF Galaxy Redshift Survey (2dFGRS),  which  is a deeper and denser  survey. Their analysis considers  $127,000$  galaxies in two regions centred near the south and north galactic poles, covering the redshift range $ 0.03 < z < 0.25$.  Considering triangle configurations in the wave-number range $0.1 < k < 0.5 \, h \, {\rm Mpc}^{-1}$, they have analysed the shape dependence of the bispectrum to measure $b_1$ and $b_2$. \citet{2004Jing} have measured the three-point correlation function of the 2dFGRS considering galaxy samples of  different luminosity.    
 \citet{2015GilMarin} have analysed the power spectrum and bispectrum for a sample of $690,827$ galaxies from  the Baryon Oscillation Spectroscopic Survey (BOSS)  CMASS Data Release 11 
 covering the redshift range $0.43 < z < 0.70$ with an effective volume of $ \sim 6  \, {\rm Gpc}^3$  and a number density  of $n \sim  3 \times  10^{-4}  \, [h\,  {\rm Mpc}^{-1}]^3$.  Their sample, which contains mostly red galaxies, is highly biased $(b_1 \sim 2)$ and hence not optimal for bispectrum studies. They find that a simple local, deterministic, Eulerian bias prescription (used in earlier works) is inadequate to model their bispectrum, and it is necessary to adopt a  non-local bias prescription. Their analysis yields estimates of three distinct combinations of the four  parameters   $b_1,b_2$, $f$ the linear growth factor and $\sigma_8$ the normalization of the linear dark matter power spectrum.  \citet{dAmico_2020} have used effective field theory to model the power spectrum and bispectrum of the 
DR12 BOSS data. However, their treatment of the bispectrum is somewhat restrictive. They find that including the bispectrum does not make a significant improvement for their estimates of the cosmological parameters,  and hence they do not include it in their final analysis.  A number of other works have also  used the bispectrum of  the BOSS DR12 data  to  determine different cosmological parameters \citep{2017MNRASGilMarin,2018Pearson,2019MNRASGualdi, 2022DAmico}. \citet{2023PhRvDIvanov} have recently estimated the higher multipole moments for  the  bispectrum of  the BOSS DR12 data and used this,  in combination with the power spectrum, to estimate cosmological parameters. \citet{2023Hahn} have measured the BOSS DR12 bispectrum to non-linear scales ($k \le 0.5 \, h \, {\rm Mpc}^{-1}$), and employed  simulation-based inference to estimate  the cosmological parameters. 

In this paper we analyse the bispectrum at  wave numbers  $k \sim (0.075-0.434) \,{\rm Mpc}^{-1}$, 
considering  the main galaxy sample \citep{2002AJStrauss} of the Sloan Digital Sky Survey\footnote{\url{https://www.sdss.org/}} (SDSS; \cite{2000AJYork}) which has measured the images and spectra of millions of galaxies over nearly one-quarter of the entire sky. The main galaxy sample, which has a median redshift of $\sim 0.1$, 
maps the nearby galaxy distribution with unprecedented accuracy. For the present analysis we have used a  volume limited sub-sample cube of size  $[296.75\, \rm Mpc]^3$ that  contains a total $16324$ galaxies. Although our sample construction  discards a large number of galaxies, 
our data cube has a  mean galaxy number density  $n =0.63 \times 10^{-3} \, {\rm Mpc}^{-3}$, 
which is around an order of magnitude larger than that in the SDSS BOSS data analysed 
in most of the recent studies of the bispectrum. Moreover,  the simple cubic  geometry and   nearly uniform galaxy selection criteria allows us to avoid  many of the complications which 
other analyses have to deal with. Further, as noted earlier, the BOSS data is highly biased, and the local biasing prescription does not provide a good model for the bispectrum. In contrast,   
the main galaxy sample has a relatively low bias  \citep{2004ApJTegmark}, and we may expect a simple prescription to  adequately   model the bispectrum.  However, note that the BOSS volume $ \sim 6  \, {\rm Gpc}^3$  is much larger than that analysed here. 

The galaxy bispectrum has been intensely studied for more than twenty-five years, primarily  as a tool to measure cosmological parameters or constrain primordial non-Gaussianity. In other words, the interest is primarily in cosmological parameter estimation, and the bispectrum is just an useful tool towards achieving this. Our perspective is somewhat different, and   the present work  is entirely  restricted to analysing  the bispectrum as a descriptive statistics to quantify the LSS.  We do not consider cosmological parameter estimation here. 
The bispectrum is the lowest order statistics which is sensitive to the non-Gaussianity of the LSS, and it is of considerable interest to quantify how the bispectrum $B(k_1,k_2,k_3)$  depends on the shape and size of the triangle  $(k_1,k_2,k_3)$. In a recent work, \citet{2020MNRASBharadwaj} have proposed a new method to parameterize the shape and size of the  triangle using the length of the largest side $k_1$  for the size,  
and two dimensionless parameters $(\mu$,$t)$  for the shape
\citep{Mazumdar:2020bkm,Mazumdar:2022ynd}. 
In this work we have utilized a  recently introduced \citep{2021JCAPAKSHAW} fast bispectrum estimator to quantify the shape and size dependence of the binned bispectrum $B(k_1,\mu,t)$ for the main galaxy sample of SDSS. 
The primary motivation here is to investigate how the bispectrum, the lowest order statistical measure sensitive to non-Gaussian features within the distribution of galaxies, changes as the configuration of triangles in $k$-space alters.
Our analysis considers a very large number  of  triangles ($\sim 1.37 \times 10^8$), and it is extremely  challenging to assimilate and  visualize the variation of the bispectrum across the space of triangle configurations. In the present work we demonstrate how it is possible to combine the information in all of these triangles, and thereby visualize the shape and size dependence by suitably binning the bispectrum in  $(k_1,\mu,t)$. We also compare our results with those for mock galaxy samples constructed from biased N-body simulations. 

The bispectrum also serves as a valuable tool for studying galaxy formation and evolution. The red and blue galaxies trace different environments \citep{Cooray2005, Pandey2020, Bhambhani2022} and have distinct formation histories \citep{Bell2003,Dekel2004}. Studying their bispectrum can reveal how various environment driven mechanisms such as mergers, interactions, or feedback processes impact galaxy properties and their evolution. Additionally, galaxy properties may also vary based on their large-scale environments due to assembly bias. Analyzing the bispectrum of red and blue galaxies may provide insights into how assembly bias shapes their clustering properties. It is further interesting to explore how the shape and size dependence of the bispectrum changes depending on various galaxy properties.
In this work, we study the bispectrum of red and blue galaxies obtained from the SDSS sample.


The plan of our paper is as follows. In Section \ref{data} we describe the Data, both  the SDSS data and the N-body simulations. We further discuss the classification into red and blue galaxies. We also outline the bias prescription, and describe how the mock galaxy samples were constructed. We briefly present the Methodology  for bispectrum estimation in Section \ref{method}. The Results are presented in  Section \ref{results},  and we finally present Summary and Discussions in Section \ref{summary}. The paper also contains an Appendix which presents  the detailed results for the shape and size dependence of $B(k_1,\mu,t)$. 

Throughout this paper we have used a  $\Lambda \rm CDM$ cosmology, with matter density parameter $\Omega_m = 0.315$, cosmological constant $\Omega_{\Lambda} = 0.685$, dimensionless Hubble parameter $h = 0.674$, normalization $\sigma_8=0.811$ and the spectral index of the primordial power spectrum $n_s = 0.965$ \citep{2020A&APlanck}.
\section{Data}
\label{data}
In this work, we use data from the SDSS to measure the galaxy bispectrum. We compare our results with the predictions obtained from a set of mock data constructed using $\Lambda$CDM N-body simulations. We have also used multiple realizations of the mock data, $50$ in number, to estimate error-bars for the measured galaxy bispectrum (and also power spectrum). 
The SDSS and the N-body data used in our work are described in this section.

\begin{figure}[htbp!]
    \begin{subfigure}{0.3\textwidth}
        \centering
        \includegraphics[width=1.5\textwidth]{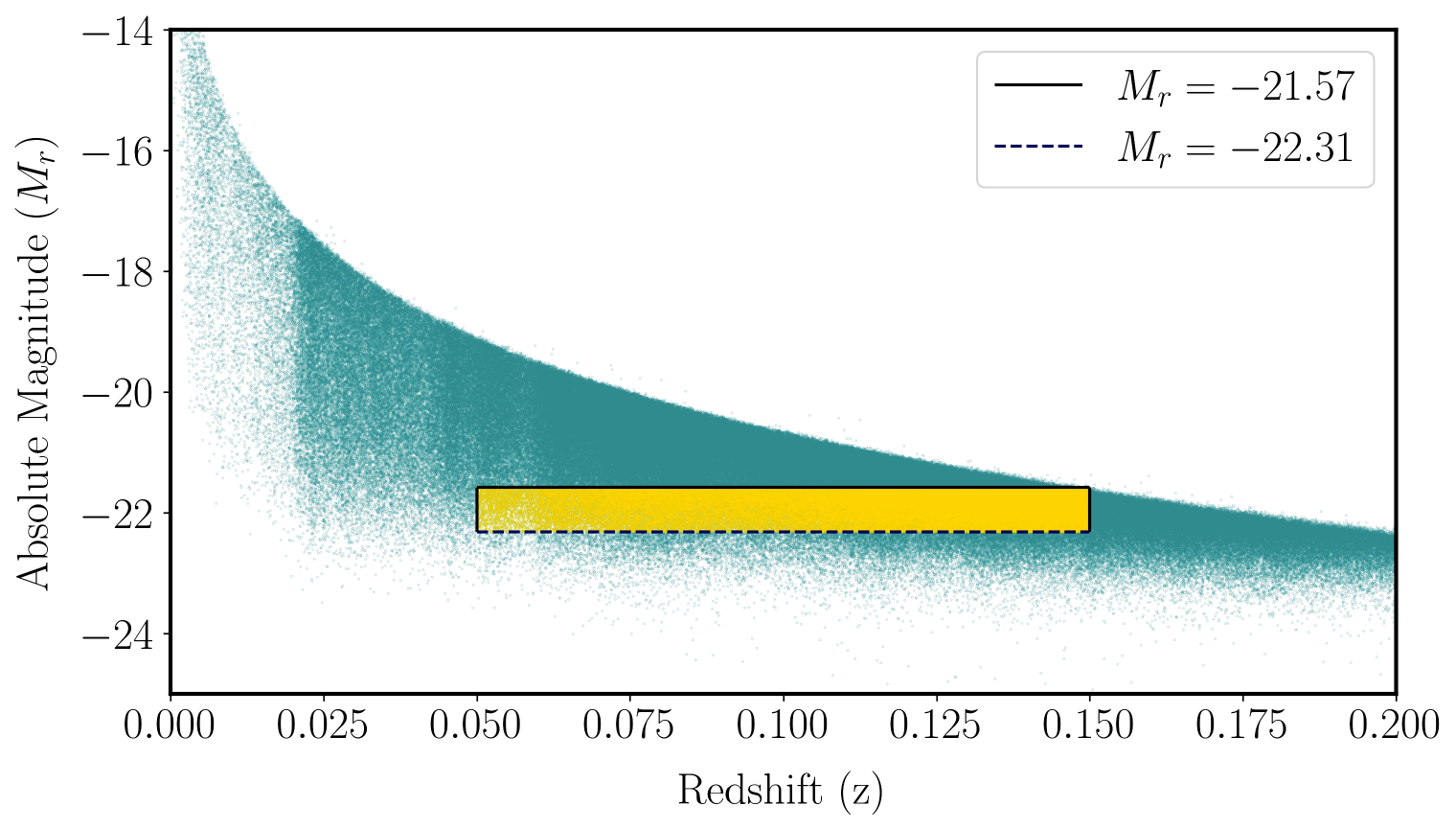}
        \caption{}
    \end{subfigure}%
    \hspace{3.8cm}
    \begin{subfigure}{0.3\textwidth}
        \centering
        \includegraphics[width=1.34\textwidth]{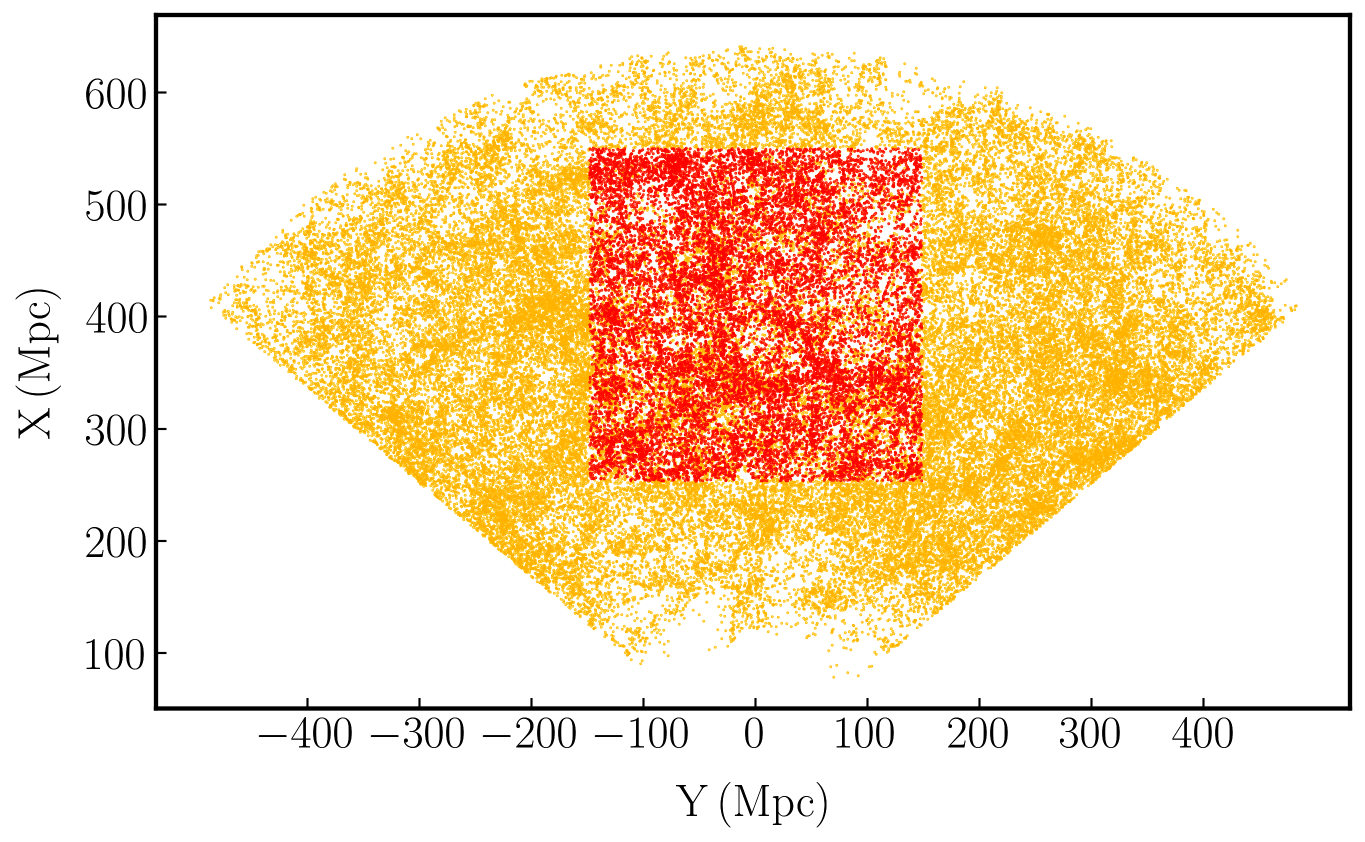}
        \caption{}
    \end{subfigure}%
    \hfill
    \vspace{0.2cm}
    \begin{minipage}{\textwidth}
        \centering
        \hspace{-1.4cm}
    \begin{subfigure}{0.4\textwidth}
        \centering
        \includegraphics[width=0.97\textwidth]{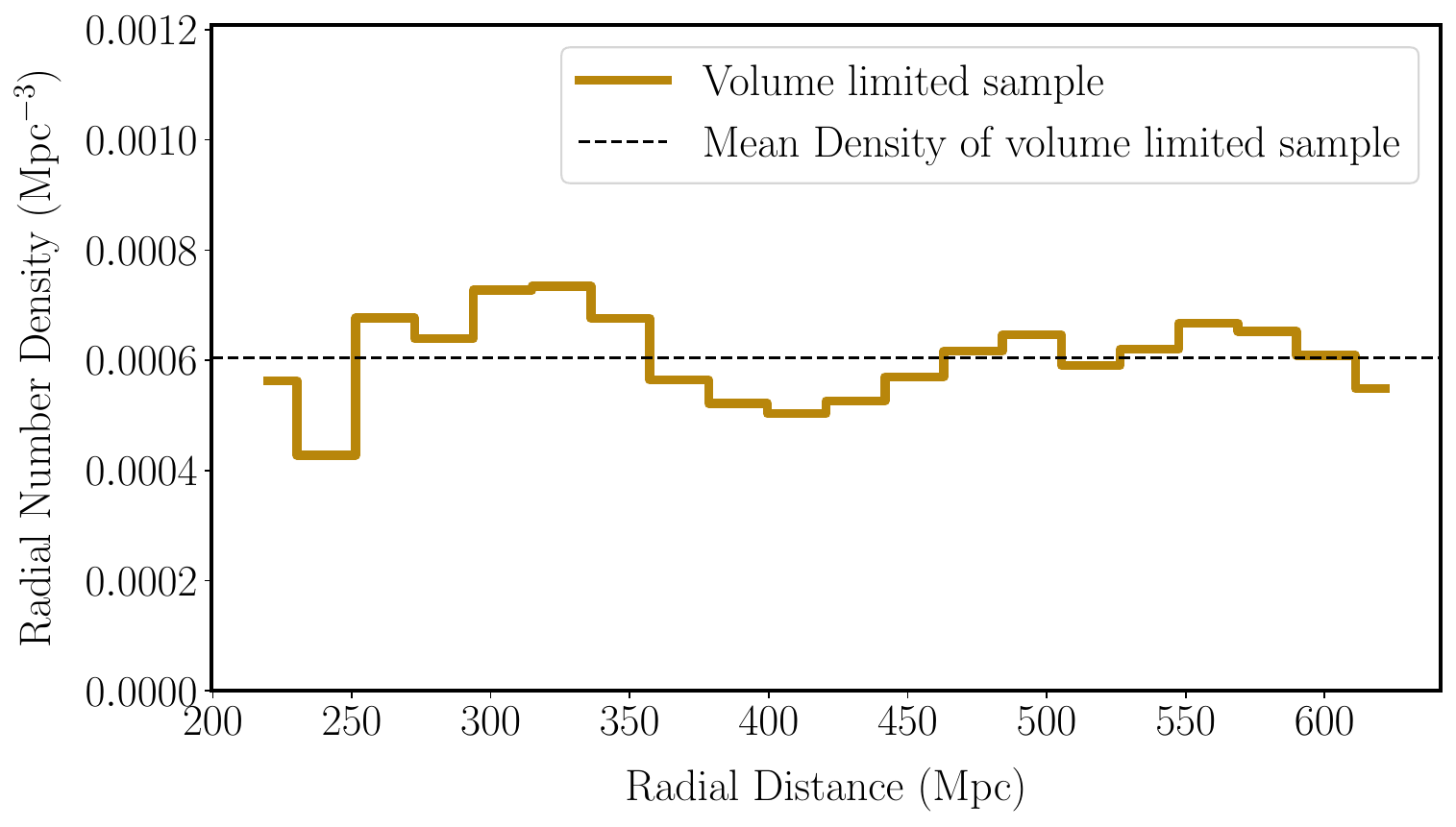}
        \caption{}
    \end{subfigure}
    \end{minipage}
    \caption{Preparation of the volume limited sample from SDSS data. In (a) we show the definition of the volume limited sample in the redshift-absolute magnitude diagram. In (b) we represent the projected distribution of our volume limited sample along with the extracted data cube that we use in the present work. The figure in the bottom panel (c) shows the radial number density of the volume limited sample along with the mean number density.}
    \label{fig:vls}
\end{figure}
\subsection{SDSS Data}
We use the data from the publicly accessible database of the SDSS. The SDSS is one of the largest redshift surveys to date. It uses a $2.5$ m telescope \citep{2006AJGunn} at Apache Point Observatory in New Mexico and provides the photometric and spectroscopic information of millions of galaxies in five photometric passbands $u, g, r, i, z$ \citep{1996AJFukugita, 1998AJGunn}. The survey covers a total $\sim 14000$ square degrees of the sky. The technical details of the SDSS are provided in \citet{2000AJYork}.

DR17 \citep{2022ApJSAbdurrouf} is the final data release of the fourth phase \citep{2017AJBlanton} of SDSS. We obtain the spectroscopic data from the main galaxy sample \citep{2002AJStrauss} using DR17, which incorporates all the targets from the prior data releases with updated and corrected measurements. The main galaxy sample consists of galaxies brighter than the $r$-band Petrosian magnitude $m_r \leq 17.77$, with a median redshift of $\sim 0.1$. We further apply a bright limit $m_r \geq 14.5$ to mitigate the effect of incompleteness due to bright galaxies.

We download the data from the \textit{Catalog Archive Server}\footnote{\url{https://skyserver.sdss.org/casjobs/}} using the \textit{Structured Query Language}. A contiguous region spanning $130^\circ \leq \alpha \leq 230^\circ$ and $0^\circ \leq \delta \leq 62^\circ$ is selected for our analysis. We prepare a volume limited sample by restricting 
the galactic extinction corrected and k-corrected $r$-band absolute magnitude $-22.31 \leq M_r \leq - 21.57$. The redshift range of the resulting sample is $0.05 \leq z \leq 0.15$ and contains a total $79235$ galaxies. We extract the largest cube that can be fitted within our volume limited sample. It has a size of $[296.75\, \rm Mpc]^3$ and contains a total $16324$ galaxies  with a median redshift of $0.102$.
 The mean galaxy number density in the data cube is $0.63 \times 10^{-3} \, {\rm Mpc}^{-3}$. The geometry and the definition of our volume-limited sample are described in Figure~\ref{fig:vls}. The radial number density of the sample is roughly the same throughout the extent of data volume.

\subsection{Classification of SDSS galaxies}
\begin{figure*}[htbp!]
    \includegraphics[width = 0.9\textwidth]{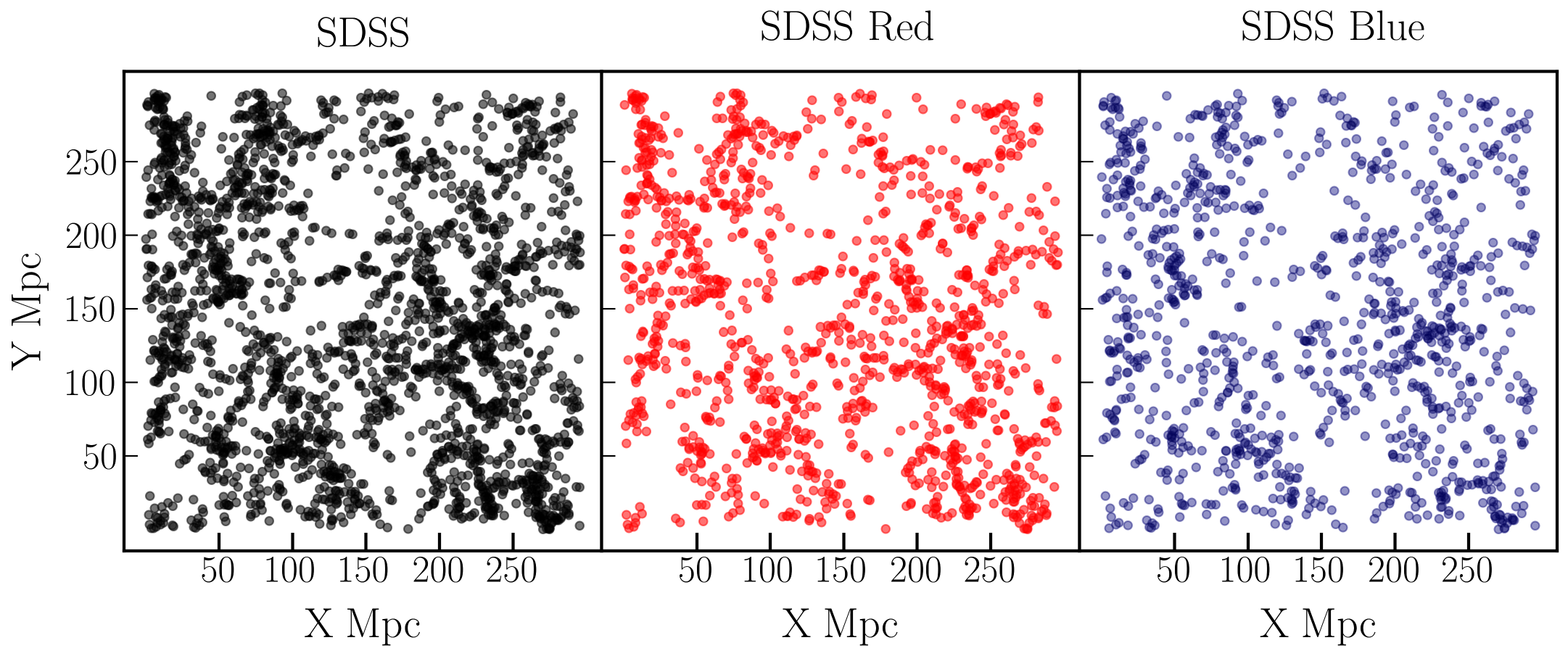}
    \caption{Two dimensional slices of galaxy distribution. The different panels show the galaxy distributions in a 2D slice of thickness $50\,\rm Mpc$ through the SDSS data cube, along with the same for red and blue galaxy samples.} 
    \label{fig:sdss_galaxies}
\end{figure*}

We further classify our SDSS galaxy sample as part of the `red sequence' and `blue cloud' depending on their distribution in (u-r) colour-stellar mass plane, using the Otsu's thresholding technique \citep{Otsu1979}. Here we provide a brief description of the classification scheme, detailed analysis can be found in \citep{Pandey2022}. This method utilizes the bimodal nature of the galaxy (u-r) colour distribution (see Figure~\ref{fig:color_pdf}). Starting with an initial colour threshold value, it splits the galaxies into red and blue classes. The galaxies having  (u-r) colour greater than the threshold value are red, while blue galaxies are those whose colour value is less than threshold. The optimum threshold is obtained by iteratingly searching for a colour value at which the within-class variance is minimum or between-class variance is maximum. However, due to the strong correlation between galaxy colour and stellar mass, applying a single colour cut to classify the entire galaxy sample is inappropriate. Consequently, we divide the sample into different stellar mass bins and apply the same methodology to determine the (u-r) colour threshold at each bin.
Figure~\ref{fig:red_blue_classification} shows the colour thresholds obtained at each mass bin and the classified galaxies in (u-r)\ colour-$\log_{10}(\frac{M_{\star}}{M_{\odot}})$ plane. A total of $8965$ galaxies are identified as red and $6621$ galaxies as blue in our sample. The mean number density is $0.34 \times 10^{-3}\,\rm Mpc^{-3}$ for red galaxies and $0.25 \times 10^{-3}\,\rm Mpc^{-3}$  for blue galaxies. Figure~\ref{fig:sdss_galaxies} illustrates the distribution of galaxies on sky-plane in a 2D slice according to their classification.

\subsection{N-body Simulations}
We use a Particle-Mesh (PM) N-body code\footnote{\url{https://github.com/rajeshmondal18/N-body}} \citep{Bharadwaj2004} to simulate the distributions of $256^3$ dark matter particles on a $512^3$ grids covering a comoving box of size  $[512\, \rm Mpc]^3$.  We generate  $50$ independent realizations of the dark matter distributions at $z = 0.102$ which is the median redshift of our galaxy sample. 

\subsubsection{Biased Realizations}
\label{sec:biased_nbody}
We use the output of our N-body simulations to prepare biased particle distributions whose bispectrum we compare with that of the galaxy distribution. Our biasing algorithm, which follows \citet{1998MNRASCole}, starts with smoothing the gridded dark matter density field with a Gaussian kernel of radius $4\,\rm Mpc$. Here we use Cloud-In-Cell (CIC) scheme to compute density on the grids from N-body particles. We then define a dimensionless variable $\nu = \frac{\delta_s(\vec{r})}{\sigma_s}$, where $\delta_s(\vec{r}) = \frac{\rho_s(\vec{r}) - \bar{\rho}}{\bar{\rho}}$ is the smoothed density contrast  and $\sigma_s^2 = \langle \delta_s^2 \rangle$.
We choose a threshold $\nu_{\rm th}$ and retain only the dark matter particles associated with the grids having $\nu > \nu_{\rm th}$.  
Note that, we have chosen a coarser grid size to compute the density contrast
compared to the original N-body simulations, and retain the particles whose locations 
fall nearest to the grids that meet the above threshold.
The resulting distribution is  biased with respect to the original dark matter distribution. Considering $P_o(k)$ and $P_b(k)$ which are the power spectra of the original and biased particle distributions respectively, we estimate the average  linear bias parameter as,  $b_{1} = \sqrt{\frac{P_b}{P_o}}$ over the  range $k = (0.2-1)\, {\rm Mpc}^{-1}$.
We have used $\nu_{\rm th} = -0.3$ and $0.01$ to prepare two sets of biased distributions having average linear bias $b_1 = 1.2$ and $1.4$, respectively. The threshold values $\nu_{\rm th}$ to obtain these biased distributions are determined by trial and error method.

\begin{figure*}[htbp!]
    \includegraphics[width = 0.9\textwidth]{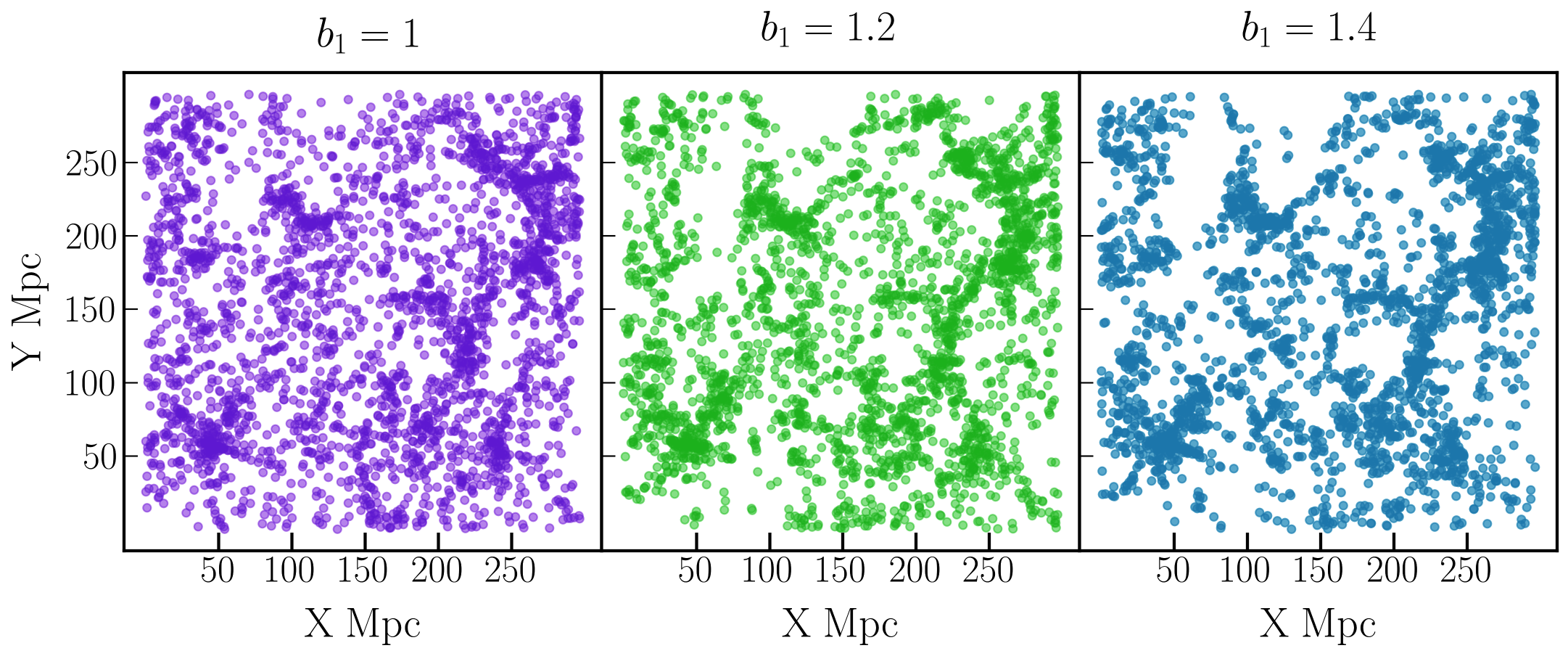}

    \caption{Similar to Figure~\ref{fig:sdss_galaxies}, but for one realisation of the mock galaxy sample for three different values of the bias as indicated in the figure.} 
    \label{fig:biased_galaxies}
\end{figure*}

\subsubsection{Preparation of Mock Samples from N-body Data}
The three-dimensional distribution of the galaxies in our SDSS data cube is obtained from the equatorial co-ordinates and the spectroscopic redshifts. In addition to the cosmological expansion, the galaxy redshifts also have a contribution from $\vec{v}_p $  
 the peculiar velocity. This distorts the galaxy clustering pattern  in redshift space. We need to map the simulated particle distributions to redshift space in order to make a fair comparison with  the SDSS results.  The  center of the SDSS data cube is located at a  distance of $428.80  \, {\rm Mpc}$  from the observer. We first place the center of the N-body cube at exactly the  same position as that of the SDSS data cube with respect to the observer, and   map the simulated particle distribution to redshift space using 
\begin{equation}
\vec{s} = \vec{r} + \frac{\vec{v}_p.\hat{{r}}}{a\,H(a)} \hat{{r}}\,,
\end{equation}
where $\vec{r}$ and $\vec{s}$ are the comoving distances in real and redshift space respectively, and $\hat{{r}}$ is the unit vector along the line of sight direction with respect to the observer.  Note that we do not use the plane-parallel approximation, and have accounted for the radial nature of the redshift-space distortion. 
Next, we  select a cubic region of  size $[296.75\,\rm Mpc]^3$, exactly the same as the SDSS volume limited sample, and randomly select  $16324$ particles to generate the mock galaxy samples to keep its number density same as the SDSS sample used. The same procedure was carried out for the biased and the unbiased particle distributions. We finally have  $50$ realizations of mock galaxy distributions corresponding to $b_1=1, 1.2$ and $1.4$ each.

The different panels of Figure~\ref{fig:biased_galaxies} show a visual representation of the galaxy distribution for a single realization of the mock galaxy samples corresponding to $b_1=1,~1.2$ and $1.4$, respectively. We note that the galaxy clustering in the  $b_1=1.2$ mock sample exhibits a  visual similarity to the clustering in the SDSS.  
However, a rigorous validation of this visual observation requires quantitative analysis which we present later in the upcoming sections. 
\section{Methodology}
\label{method}
Considering the galaxy distribution in our SDSS (or N-body) sample, we have used the CIC algorithm to calculate the galaxy density contrast  $\delta_g(\vec{x}) $  on a $256^3$ grid  
and used Fast Fourier Transform (FFT) to calculate its Fourier conjugate $\Delta_g(\vec{k})$. 
The galaxy power spectrum $P_g(k)$ and bispectrum $B_g(k_1,k_2,k_3)$ are respectively defined as, 
\begin{equation}
  P_g(k) = V^{-1} \langle \Delta_g(\vec{k})\Delta_g(-\vec{k})\rangle \, ,
\end{equation}
\begin{equation}\label{eq:bs_def}
   B_g(k_1,k_2,k_3) = V^{-1} \langle \Delta_g(\vec{k}_1)\Delta_g(\vec{k}_2)\Delta_g(\vec{k}_3)\rangle \,, 
\end{equation}
where $\vec{k}_1, \vec{k}_2$ and $\vec{k}_3$ form a closed triangle $(\vec{k}_1+\vec{k}_2 + \vec{k}_3 = 0)$.
  Considering the discrete galaxies as a Poisson sampling of an underlying smooth matter density field, it is necessary to apply  a shot noise correction \citep{1980Peebles} . The shot noise corrected galaxy power spectrum and bispectrum are respectively computed using  
\begin{equation}
    P(k) = P_g(k) - \bar{n}_g^{-1} \, 
    \label{eq:ps1}
\end{equation}
and 
\begin{equation}
    B(k_1,k_2,k_3) = B_g(k_1,k_2,k_3) - {\bar{n}_g}^{-1}  {[P(k_1) + P(k_2) + P(k_3)]} -{\bar{n}_g^{-2}} \, ,
    \label{eq:bs1}
\end{equation}
where $\bar{n}_g$ is the mean galaxy number density.

In the presence of redshift-space distortions, the value of the bispectrum depends on how the triangle $(\vec{k}_1, \vec{k}_2,\vec{k}_3)$ is oriented with respect to the line-of-sight direction. However, the spherically averaged bispectrum $B(k_1,k_2,k_3)$ for which   results are presented here has been averaged over all triangle orientations, and it depends only  on the shape and size of the triangle $(k_1,k_2,k_3)$. Here we assume that the three sides are ordered such that $k_1 \ge k_2 \ge k_3$. 
Following \citet{2020MNRASBharadwaj}, we have used  $k_1$ and $(\mu,t)$ to respectively 
 parameterize the size and shape of a triangle. Here $k_1$ refers to the largest side, 
\begin{equation}
    \mu = (2 k_1 k_2)^{-1} \, [k_1^2 + k_2^2 - k_3^2] \, 
    \label{eq:mu}
\end{equation}
   is the cosine of the angle between $k_1$ and $k_2$, and 
\begin{equation}
    t=k_2/k_1
    \label{eq:t}
\end{equation}   
    is the ratio of the second largest side to the largest side.  The values of $(\mu,t)$ are further restricted to the range  $ 0.5 \leq t,\mu \leq 1$   and $2t\mu \geq 1$.

For the present work, we have used the fast binned bispectrum estimator presented in \citet{2021JCAPAKSHAW} to estimate $B_g(k_1,\mu,t)$. 
This estimator is carefully crafted to retain only the $k$-modes that do not have the aliasing effects.
Our estimator first divides the 3D $\vec{k}$-space of interest into a number of linearly-spaced spherical shells of single grid spacing $\delta k=0.021 \, {\rm Mpc}^{-1}$. 
Considering three such shells labelled ${\rm S}_1$, ${\rm S}_2$  and ${\rm S}_3$ with mean radii $k_1 \ge k_2 \ge k_3$  respectively, the estimator provides the binned bispectrum averaged over all closed triangles which have one side in each of the shells. We have used $(k_1,k_2,k_3)$ in equations~(\ref{eq:mu}) and (\ref{eq:t})  to calculate   $ (k_1,\mu,t)$  corresponding to this particular  bin of triangles. We subsequently apply the shot noise correction (equation~\ref{eq:bs1})  to obtain the bispectrum $B(k_1,\mu,t)$. 
We find that for large $k_1$ the shot noise correction dominates the estimated bispectrum, and we have excluded these estimates from the subsequent analysis. Considering all possible combinations of three shells,  we find that the resulting distribution of binned bispectrum estimates are not uniformly distributed across the $(\mu,t)$  plane \citep[see Figure 2 of][]{2021JCAPAKSHAW}. To achieve a more uniform coverage in the $(\mu,t)$ plane,  and also increase the signal to noise ratio we have further binned the bispectrum estimates. For every $k_1$, we have divided the $(\mu,t)$ plane into $5\times 5$ uniform cells and averaged the estimates which fall within each cell, weighing each estimate with the corresponding number of triangles.  We finally have estimates of the bispectrum  $B(k_1,\mu,t)$ for twenty $k_1$ values spanning $(0.034-0.434) \, {\rm Mpc}^{-1}$ with $5 \times 5$ bins in $(\mu,t)$ plane for each $k_1$. However, we find that the entire $(\mu,t)$ plane is not populated for small $k_1$, or in other words the $k_1$ range where we have estimates of the bispectrum varies with the shape of the triangle $(\mu,t)$.

The SDSS data and the mock galaxy samples both are analysed in exactly the same way. SDSS uses fibre optics to collect light from galaxies. The minimum distance between the centres of the SDSS fibres is $55^{\prime \prime}$ \citep{strauss02}. The finite size of the fibres leads to a spectroscopic incompleteness where the galaxies closer than this minimum separation are often missed. This effect is often referred to as fibre collision. We estimate that the fibre collision effect is less than $1\%$ for the SDSS galaxy sample used in this work, and ignored its impact in our analysis. For the mock samples, we have used $50$ realizations to estimate the mean and standard deviation for both the power spectrum and the bispectrum. 

\section{Results}
\label{results}
Here we present results for the power spectrum and bispectrum measured for the SDSS data as well as the various mock samples prepared from the  N-body simulations. We primarily concentrate  on the bispectrum which is the main focus of this paper. The results are also presented  for the power spectrum  which is needed as an input to estimate the bispectrum using equation~(\ref{eq:bs1}). 
\subsection{Power spectrum}
\label{res:powsp}

\begin{figure}[htbp!]
    \centering
    \begin{subfigure}[b]{0.5\textwidth}
        \centering
        \includegraphics[width=0.9\linewidth]{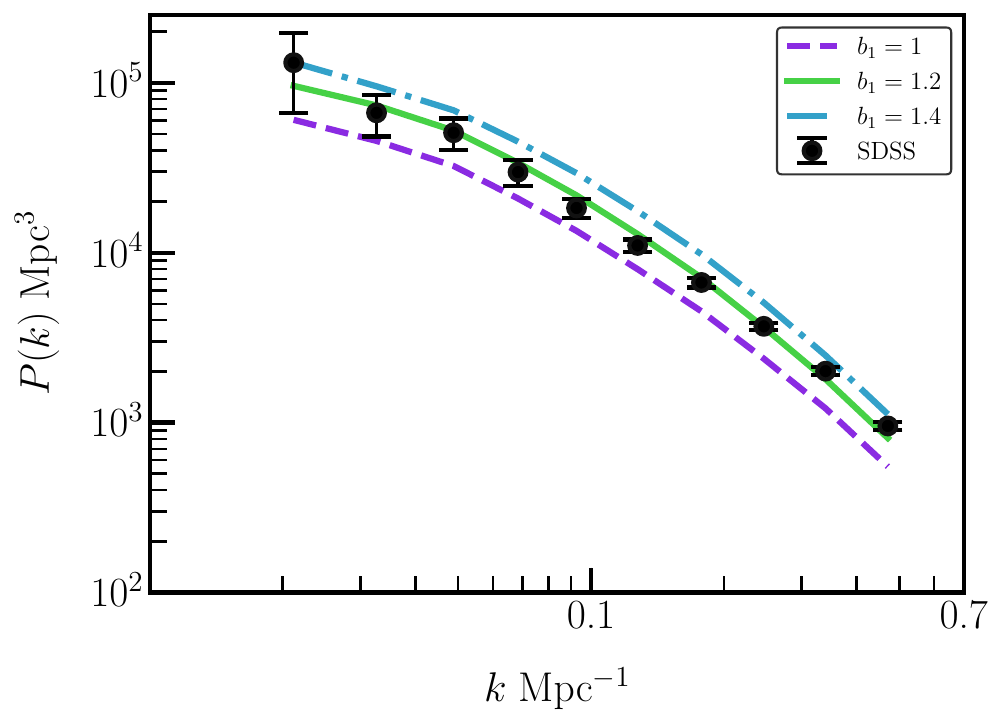}
        \caption{}
    \end{subfigure}%
    \begin{subfigure}[b]{0.5\textwidth}
        \centering
        \includegraphics[width=0.89\linewidth]{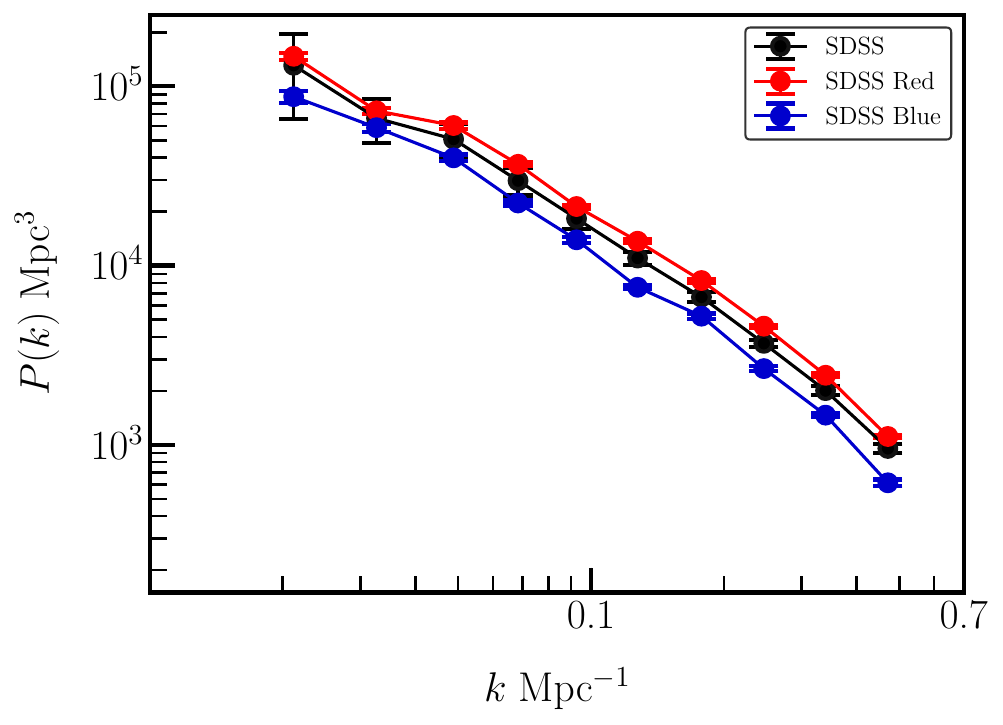}
        \caption{}
    \end{subfigure}
    \caption{Power spectrum as a function of $k$. The data points (black circles) show $\pks$ the power spectrum values for our SDSS galaxy sample. The right panel (a) shows the power spectra for three different mock samples, represented by three lines of different colors. The $1 \sigma$ error bars shown for $\pks$ have been estimated using $50$ realizations of the mock sample with $b_1 = 1.2$. (b) represents the comparison of the red and blue galaxy power spectrum with that of the entire sample.}
    \label{powsp}
\end{figure}
The black circles in Figure~\ref{powsp}  show the power spectrum $\pks$ estimated from our SDSS data. We have estimates of $\pks$ across   the $k$ range $0.021 \, {\rm Mpc}^{-1}$ to $0.471 \, {\rm Mpc}^{-1}$ which is well below the Nyquist cut-off value $2.95\, {\rm Mpc}^{-1}$.  We have also shown  $[P(k)]_{b_1}$ estimated from the different mock samples corresponding to $b_1=1,~1.2$ and $1.4$ respectively. We see that $[P(k)]_{1.2}$, which is the result for $b_1=1.2$, is in close agreement with the $\pks$.  Our mock galaxy samples incorporate all the known observational effects like finite volume, number density and RSD which influence the actual data, and we have used the mock samples to predict the statistical uncertainty for the
 estimated $\pks$.  In the present analysis, we have adopted the $1 \sigma$ error estimated from $50$ realizations of the $b_1=1.2$ mock samples to be the predicted statistical uncertainty for $\pks$, and these are shown as $1 \sigma$ error bars for the data points in the left panel of Figure~\ref{powsp}. Further, we have interpolated the values of the estimated $\pks$  and used these in  equation~(\ref{eq:bs1}) to estimate the SDSS bispectrum. We have similarly interpolated $[P(k)]_{b_1}$ for the different mock samples and used these to estimate the corresponding bispectrum presented in the next section. We note that, due to this interpolation, the binning effects of the power spectrum does not impact the bispectrum estimates significantly, and it remains within $1\sigma$ error.

Although a visual comparison (left panel of Figure~\ref{powsp}) indicates that  $\pks$ is in good agreement with $[P(k)]_{1.2}$,   a quantitative comparison reveals a reduced chi-square of $\bar{\chi^2}=1.89$ for which the $p$ value is $0.026$. 
To test whether this mismatch can be alleviated by considering a slightly higher value of the linear bias $b_1$, we have scaled the $[P(k)]_{1.2}$ to define $[P(k)]_{b_1}=(b_1/1.2)^2 \times [P(k)]_{1.2}$ for other values of $b_1$ in the range $1.2 \le b_1 \le 1.4$. We find that the best match is obtained at $b_1=1.21$ for which we have $\bar{\chi^2}=1.66$ and $p=0.08$ that is acceptable.  We do not expect the error bars to change much if we use $[P(k)]_{b_1}$ with  $b_1=1.21$ instead of $b_1=1.2$.  Hence, for the present analysis, we have used $[P(k)]_{1.2}$ to estimate the error bars for $\pks$.

\subsection{Bispectrum}

\begin{figure*}[htbp!]
    \includegraphics[width = 1.\textwidth]{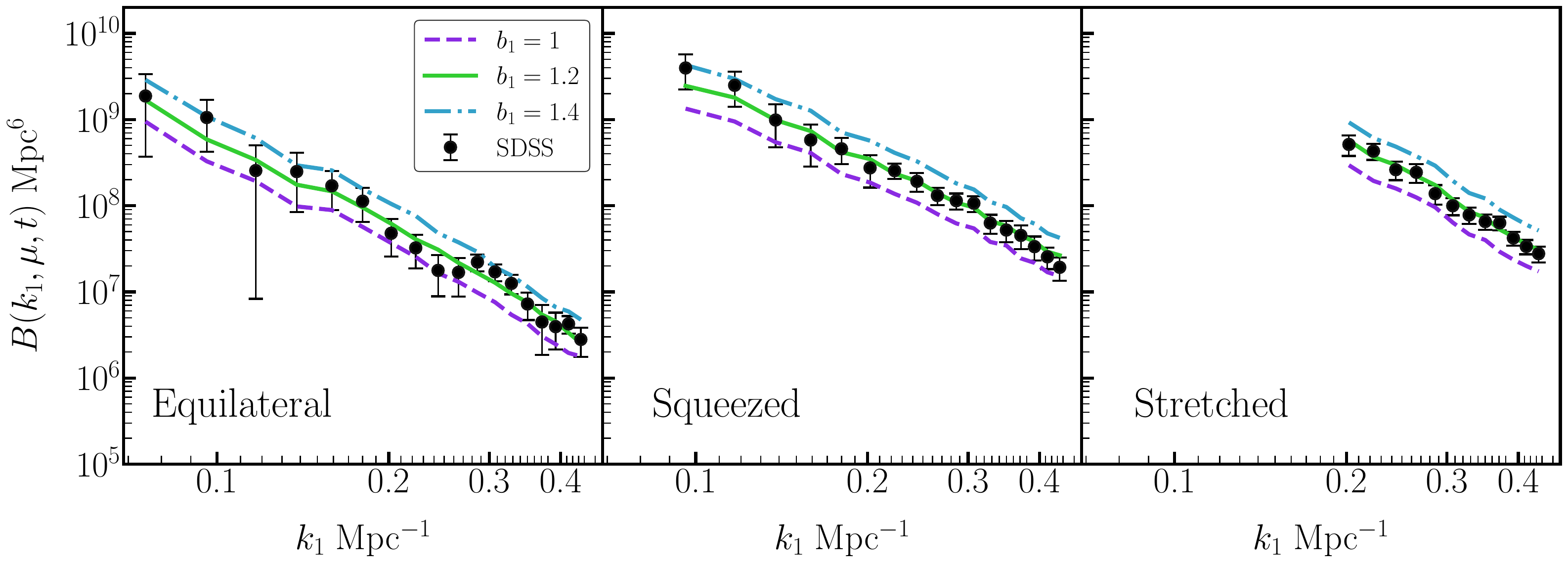}
    \caption{Bispectrum $\bks$ as a function of $k_1$. This figure shows the SDSS bispectrum (black circles) for three different triangle shapes namely equilateral, squeezed and stretched in the three different panels. The results from mock galaxy samples with bias $b_1=1,~1.2$ and $1.4$ are also shown for comparison. The error bars are the $1 \sigma$ errors from $b_1=1.2$ mock sample which provides a good match with the SDSS results.}
    \label{B_vs_k}
\end{figure*}
\begin{figure*}[htbp!]
    \hspace{1cm}
    \includegraphics[width = .94\textwidth]{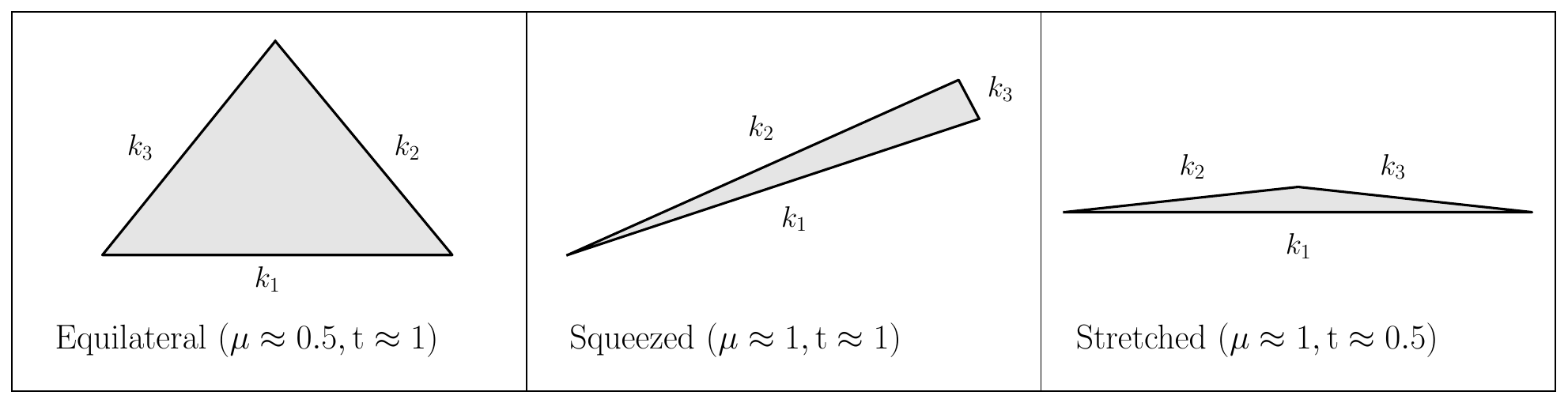}
    \caption{Traingle configurations. This plot illustrates equilateral, squeezed, and stretched triangle shapes in the left, center and right panels, respectively.}
    \label{fig:three_triangles}
\end{figure*}

We now analyse the  binned bispectrum $\bks$ estimated from the SDSS data. As discussed in Section~\ref{method}, we have considered twenty $k_1$ shells in the range  $(0.034-0.434) \, {\rm Mpc}^{-1}$. However,  we have very few triangle shapes  $(\mu,t)$ at  small $k_1$.  The number of triangles in each bin is also very small,  leading to large statistical errors in $\bks$. As a consequence, we have discarded some of the small $k_1$, and the final $k_1$ range varies with the triangle shape. We note that  the $k_1$ values span the largest range $(0.075 - 0.434) \,\rm Mpc^{-1}$, ($18$ shells) for equilateral triangles.

Figure~\ref{B_vs_k} shows  the $k_1$ dependence of $\bks$ for three specific triangle shapes namely equilateral, squeezed and stretched which respectively correspond to $(\mu,t) \approx (0.5,1), (1,1)$ and $(1,0.5) $ .  These three triangle shapes are explicitly illustrated in Figure~\ref{fig:three_triangles}. We find  very similar $k_1$ dependence for all three shapes where $\bks$ appears to be a power law 
\begin{equation}
    B(k_1,\mu,t)=A \big(k_1/1\mpci\big)^{n}
    \label{eq:fit}
\end{equation}
with a negative index $n \, (< 0)$.  The amplitude $A(\mu,t)$ and index $n(\mu,t)$ both change with shape,  however to maintain brevity of notation we do not explicitly denote this $(\mu,t)$ dependence.  For comparison, in Figure~\ref{B_vs_k} we have also shown the binned bispectrum $[B(k_1,\mu,t)]_{b_1}$ estimated from the mock data for different values  of $b_1$. We see that  $[B(k_1,\mu,t)]_{b_1}$ also appears to have a power law dependence on $k_1$, very similar to $\bks$. For a fixed shape, the index $n$ does not change if $b_1$ is varied, the amplitude $A$ however increases with $b_1$. For all the three shapes,  we  see that  $[B(k_1,\mu,t)]_{1.2}$ is in reasonably good agreement with $\bks$ and we adopt the $1 \sigma$ errors for   $[B(k_1,\mu,t)]_{1.2}$ as the errors for $\bks$. Thus, unless stated otherwise, the error bars correspond to the estimates of the cosmic variance derived from $50$ realizations throughout this work. We also employed two other ways to estimate the error bars from the SDSS galaxy sample itself. In the first method, we divide the data volume into eight equal sub-volumes and compute the bispectrum for each sub-volume. 
The errors obtained using in this way have similar order of magnitude as $1 \sigma$ errors for $[B(k_1,\mu,t)]_{1.2}$. We also note that the effect due to super-sample variance is negligible, and it remains within the error bars of summary statistics. The errors are expected to decrease as the volume gets larger \citep{Mazumdar:2022ynd}, and they will be eight times smaller for full data volume. However, we do not use the errors obtained using this scheme as the $k_1$ range does not coincide with the modes being probed in this work. In another method, we prepare ten sub-samples from SDSS data by randomly selecting $80\%$ of total galaxies in each sub-sample. The error bars in this case are smaller than in previous cases. We use the $\chi^2$ statistics to quantify the match between $\bks$ and $[B(k_1,\mu,t)]_{1.2}$. We find that the reduced chi-square has values $\bar{\chi^2}= 0.51, 0.28$  and $0.53$ for equilateral, squeezed and stretched triangles respectively. We note that the $\bar{\chi^2}$ is considerably smaller than unity for the squeezed triangle case, which corresponds to overfitting. The reason may be that the errors are correlated. We further use the covariance matrix to compute $\bar{\chi^2}$. The 50 realizations may not be optimal for a robust estimate of the covariance matrix, but it serves the purpose of providing rough estimates of $\bar{\chi^2}$. After incorporating the full covariance matrix, the $\bar{\chi^2}= 0.67, 0.93$  and $1.39$ for equilateral, squeezed and stretched triangles respectively.
The best-fit estimates of A and n computed using the full covariance matrix do not vary significantly, and the values remain within quoted error bars. We use the diagonal of the covariance matrix for model fitting in the rest of the paper. It is visually apparent that $b_1=1$ and $b_1=1.4$ respectively under-predict and over-predict $\bks$, and  we  find that $\bar{\chi^2}>1$ for all three shapes. 

\renewcommand{\arraystretch}{1.5} 

\begin{table}[htbp!]
    \centering
    \begin{tabular}{|c|c|c|c|}
    \hline
     Triangle Shape & $A~({\rm Mpc}^6)$ & $n$ &   $\bar{\chi^2}$ \\
     \hline
     Equilateral & $(1.65 \pm 0.51)\times 10^5$ & $(-3.64 \pm 0.24)$ & $0.53$ \\
     \hline
     Squeezed & $(1.40 \pm 0.36)\times 10^6$ & $(-3.42 \pm 0.19)$ & $0.27$ \\
     \hline
     Stretched & $(1.47 \pm 0.41)\times 10^6$ & $(-3.62 \pm 0.24)$ & $0.33$ \\
     \hline
     
    \end{tabular}
    \caption{Fitted parameters $A$ and $n$. This shows the fitting parameters of a single power law of the form $A\,\big(k_1/1\mpci\big)^{n}$, used to fit SDSS bispectrum for three triangle configurations. The reduced chi-square $\bar{\chi^2}$ of the fitting is also shown in the right-most column.}
    \label{tab:SDSS_fit}
\end{table}

\begin{figure*}
    \includegraphics[width = 1\textwidth]{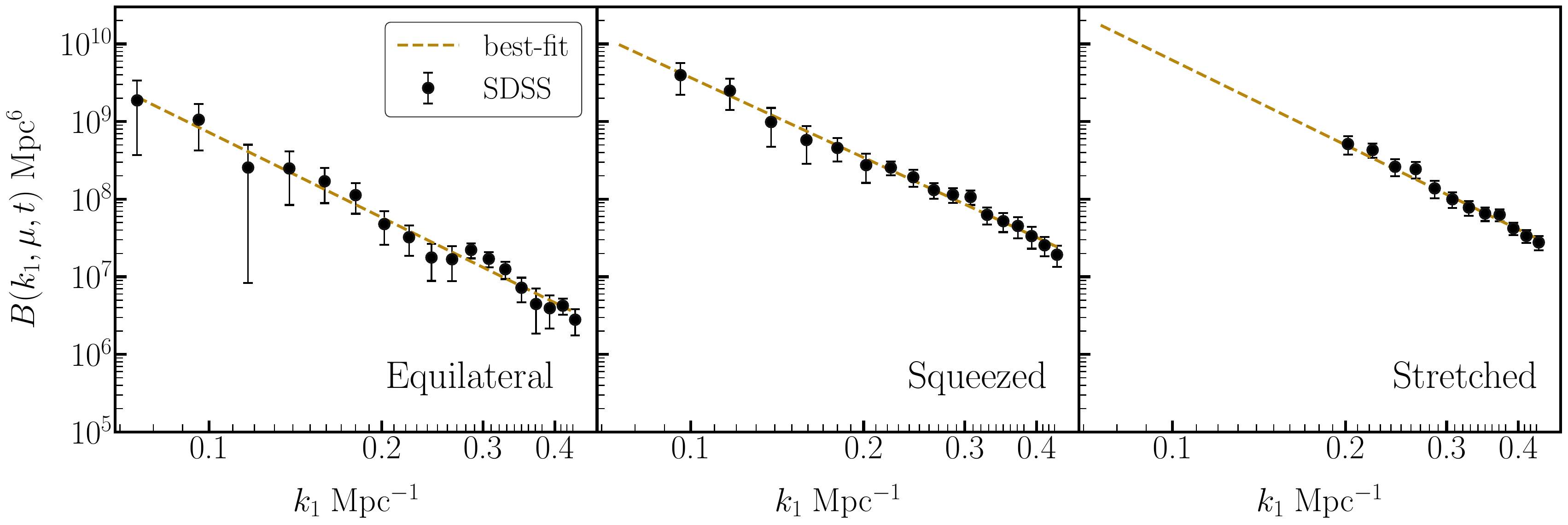}
    \caption{SDSS bispectrum and its fit as a function of $k_1$.
    We show the measured SDSS bispectra for equilateral, squeezed and stretched triangles in the three panels as labelled. We also fit the SDSS bispectra using a power law of the form $A\,\big(k_1/1\mpci\big)^{n}$. The best-fit bispectra are represented by the dashed yellow line. The $1\sigma$ error bars on the data are estimated from N-body($b_1 = 1.2$) samples.}
    \label{fig:k1_dep}
\end{figure*}


To quantitatively model the $k_1$ dependence of $\bks$, we have fitted a power law (equation~\ref{eq:fit}) for the three shapes considered in Figure~\ref{B_vs_k}. The best-fit parameters  and the $\bar{\chi^2}$ values  are tabulated in Table~\ref{tab:SDSS_fit}, while the corresponding best-fit power laws  are shown in Figure~\ref{fig:k1_dep}. 
For all  the three triangle shapes considered here,  we see that the  $k_1$ dependence of $\bks$ is well fitted by a power law with $\bar{\chi^2} < 1$ ($\bar{\chi^2}$ for all cases increases if the full covariance matrix is used). Considering the best-fit parameter values  in Table~\ref{tab:SDSS_fit}, we find that the parameter $A$, which 
represents the amplitude of $\bks$  at $k_1=1\mpci$, is maximum for the stretched triangles, followed by squeezed triangles, and $A$ is minimum for equilateral triangles. 
The values of $A$ are comparable for squeezed and stretched triangles, and these are almost an order of magnitude larger than the value for equilateral triangles. Among these three configurations, the mean value of $| n |$ is maximum for ($n \approx - 3.64$) corresponding to equilateral triangles and smallest ($n \approx - 3.42$) for squeezed triangles.


\begin{figure}[htbp!]
    \begin{subfigure}{0.5\textwidth}
        \centering
        \includegraphics[width=\linewidth]{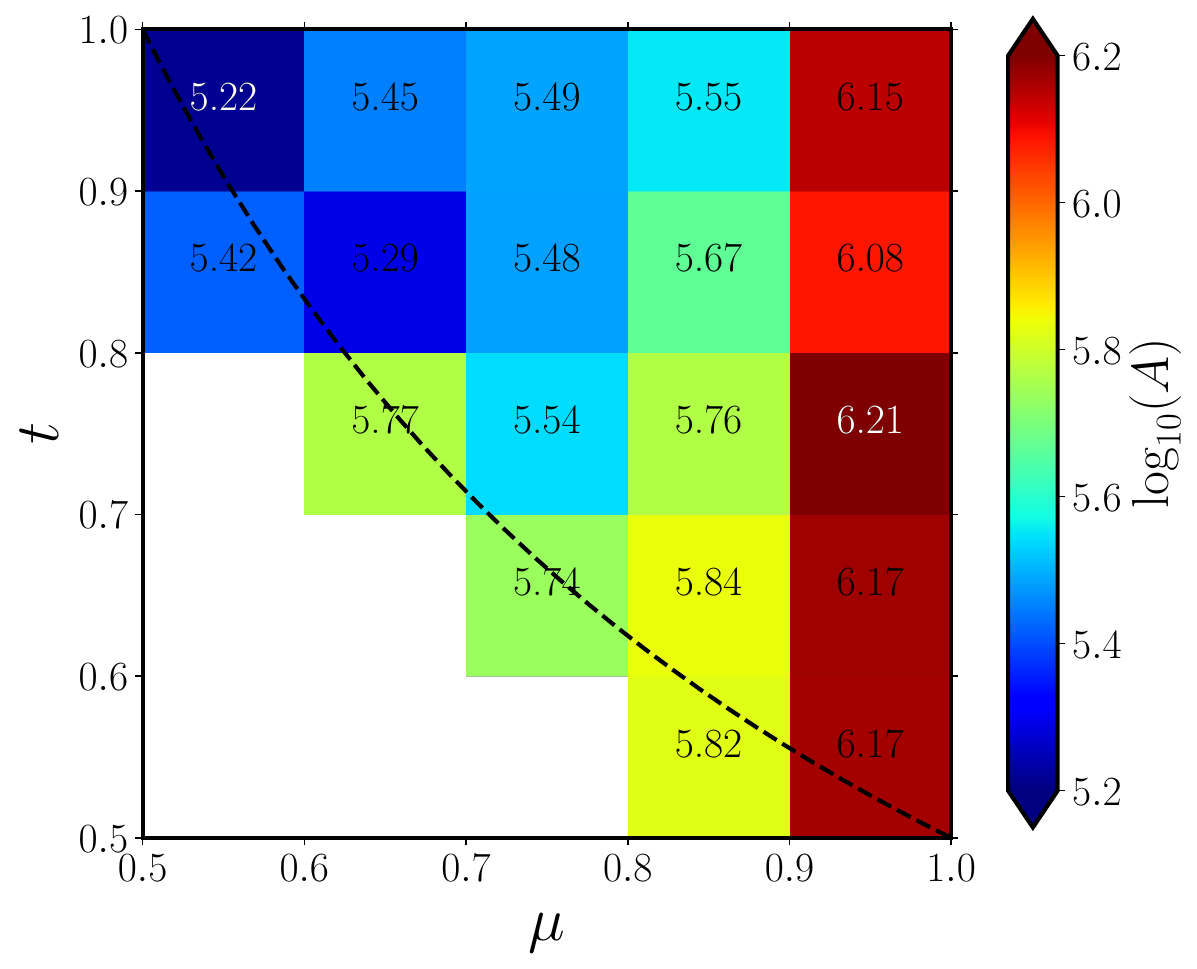}
    \end{subfigure}%
    \hspace{0.04\textwidth}
    \begin{subfigure}{0.5\textwidth}
        \centering
        \includegraphics[width=\linewidth]{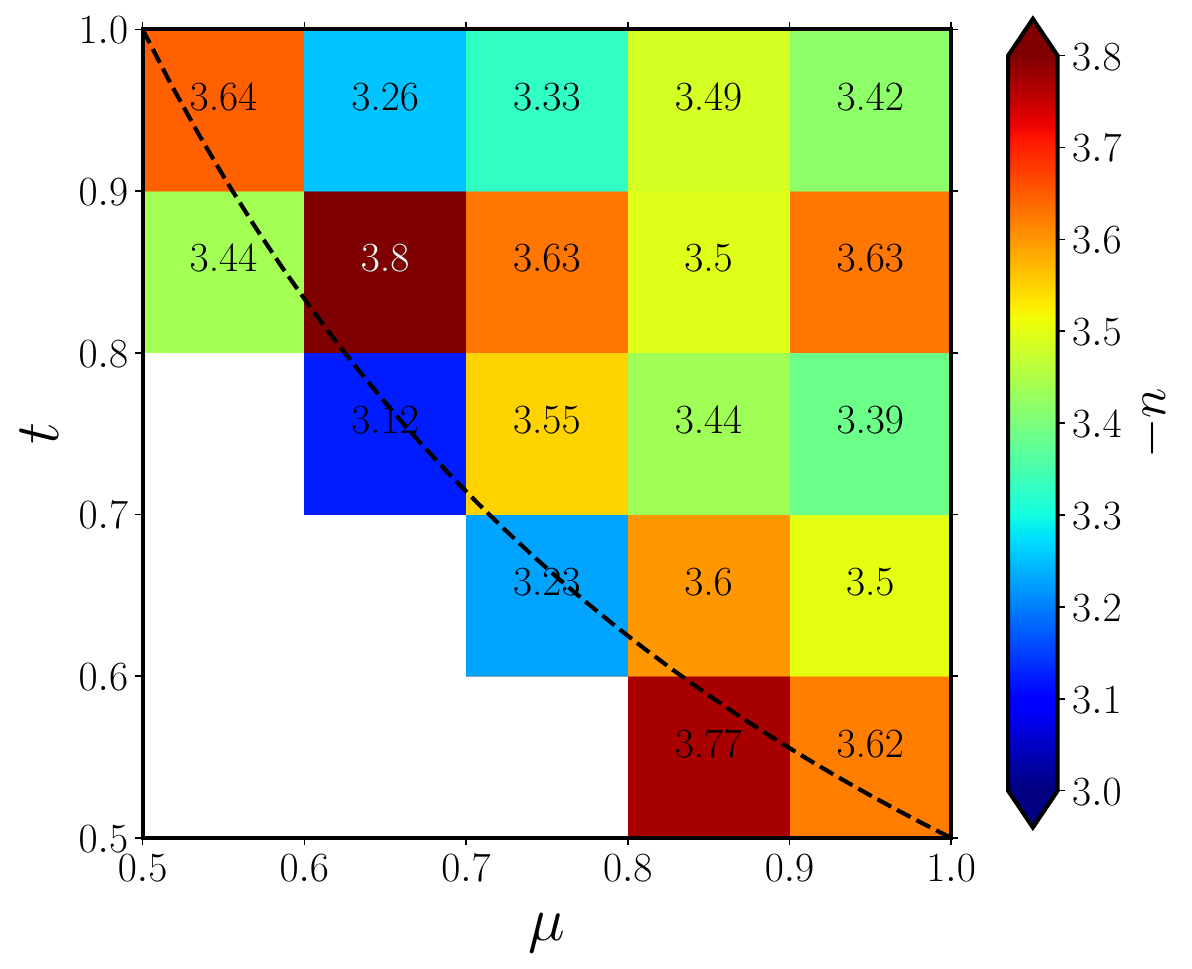}
    \end{subfigure}
\caption{The best-fit parameter values of $A$ and $n$. The left and right panels show the variations of $\log_{10}{(A)}$ and $-n$, respectively with the triangle shape parameters $\mu$ and $t$.}
\label{fig:best_A_and_n}
\end{figure}

We now consider triangles of all possible shapes as parameterized through $\mu$ and $t$  which have been defined in equations~(\ref{eq:mu}) and (\ref{eq:t}) respectively.  As mentioned in Section~\ref{method}, we have divided the region $0.5 \le \mu,t \le 1$ of the $(\mu,t)$ plane into  $5 \times 5$ cells.  Note that we have estimates of $\bks$ only in the allowed  region $2 \mu t \ge  1$, as depicted in Figure~\ref{fig:best_A_and_n}. Here  $t=1$ the upper boundary corresponds to L-isosceles triangles $(k_1=k_2)$  where  the two larger sides  are equal, $2 \mu t =1$ the lower boundary corresponds to  S-isosceles triangles  $(k_2=k_3) $ where the two smaller sides are equal, and $\mu=1$ the right boundary corresponds to linear triangles where the three sides are aligned along the same axis. The line $\mu=t$ corresponds to right-angled triangles, whereas the regions $t > \mu$ and $t<\mu$ correspond to acute and obtuse triangles respectively. The top left  corner   $(\mu,t)=(0.5,1)$ corresponds to equilateral triangles where  $k_1=k_2=k_3$. The top right corner $(1,1)$ corresponds to squeezed triangles where  
$k_1=k_2$ and $ k_3 \rightarrow 0$, whereas the bottom right corner corresponds to stretched triangles where $k_2=k_3=k_1/2$.  Note that the  three shapes for which the results have been shown in Figures~\ref{B_vs_k} and  \ref{fig:k1_dep} correspond to the three corners of the $(\mu,t)$ plane, and we see that the equilateral  triangle is slowly deformed to  a  linear triangle as we move from left to right in the $(\mu,t)$ plane. 

We have individually considered the $k_1$ dependence of $\bks$ for each of the nineteen $(\mu,t)$ bins shown in Figure~\ref{fig:best_A_and_n}, the results for which are shown in  Figures~\ref{fig:SDSS_bias_all_shapes}, \ref{fig:SDSS_ntri_all_shapes}  and \ref{fig:SDSS_bias_all_shapes_fit} which are contained in  the Appendix.  Figure~\ref{fig:SDSS_bias_all_shapes} is very similar to  Figure~\ref{B_vs_k}, except that it considers nineteen  different  $(\mu,t)$ bins which cover  the space of all possible triangle shapes uniquely. The different panels of Figure~\ref{fig:SDSS_ntri_all_shapes} show $N_{tr}(k_1,\mu,t)$ the number of triangles for each $(k_1,\mu,t)$ bin considered here. We see that the values of 
$N_{tr}(k_1,\mu,t)$ span a wide range of values from $\sim 2 \times 10^3$ to $4 \times 10^6$ 
across the different $(k_1,\mu,t)$ bins.  We also see that for any fixed $(\mu,t)$ bin, 
$N_{tr}(k_1,\mu,t)$ increases steeply (varying in the range $\sim k_1^{3}$ to $\sim k_1^{6.6}$) as $k_1$ is increased. As mentioned earlier, the total number of triangles included in our analysis is $N_{tr}({\rm Total}) \sim 1.37 \times 10^8$. Further, we also note that the main contribution to $N_{tr}({\rm Total})$ comes from the largest $k_1$ bins. 

In addition to $\bks$, each panel of Figure~\ref{fig:SDSS_bias_all_shapes} also shows $[B(k_1,\mu,t)]_{b_1}$. We find that $b_1=1.2$ closely matches  $\bks$ for every shape considered here. Considering the entire ($\mu,t)$ plane, we have $\bar{\chi^2}=1.6$ for $b_1=1.2$, whereas $\bar{\chi^2}>2$ for $b_1=1$ and $1.4$. Figure~\ref{fig:SDSS_bias_all_shapes_fit}, which is very similar to Figure~\ref{fig:k1_dep}, shows the best-fit power law (equation~\ref{eq:fit}) for all the nineteen ($\mu,t)$ bins considered here. In all cases, we find that the $k_1$ dependence of $\bks$ is well modelled by a power law with $\bar{\chi^2} < 1$ for most of the shape bins. 

The left and right panels  of Figure~\ref{fig:best_A_and_n} respectively show  how the amplitude $A$ and index $n$ of the power-law fit vary with the shape of the triangles. Considering the left panel, We find that $A$ is maximum $(\log_{10} A=6.21)$ for a linear triangle in the shape bin $\mu = 0.95,\, t = 0.75$. The value of $A$ declines rapidly as we move left (reduce $\mu$ with $t=1$ fixed) with $(\log_{10} A=5.22)$ for equilateral triangles. Overall, we have relatively large values of $A$ for linear triangles $(\mu=1)$, and the values of $A$ decline rapidly as we move left (reduce the value of $\mu$).


Considering the right panel of Figure~\ref{fig:best_A_and_n}, we see that the values of $|n|$ are all negative. We find that the magnitude of the power-law index $\mid n\mid$ is minimum $(n=-3.12)$  for the bin $\mu = 0.65,\, t = 0.75$.  The value of $| n |$ is  maximum ($n \approx - 3.8$), and we have the steepest power law  
for the triangles corresponding to the $\mu = 0.65,\, t = 0.85$ bin.

Considering the linear triangles where the bispectrum has the highest values, we find that the amplitude $A$ does not change much (less than an order of magnitude) 
as we go from $t =0.5$ to $t=1$, however the index $n$ changes considerably in this range. This means that even though we might observe the largest amplitude $A$ for the linear triangles  $\mu = 0.95,\, t = 0.75$, the maximum value of the bispectrum could occur at some other linear triangle $(\mu=1)$ configuration with $t<1$.

\subsection{Bispectrum of red and blue galaxies}
\begin{figure}[htbp!]
    \centering
    \includegraphics[width=1.\linewidth]{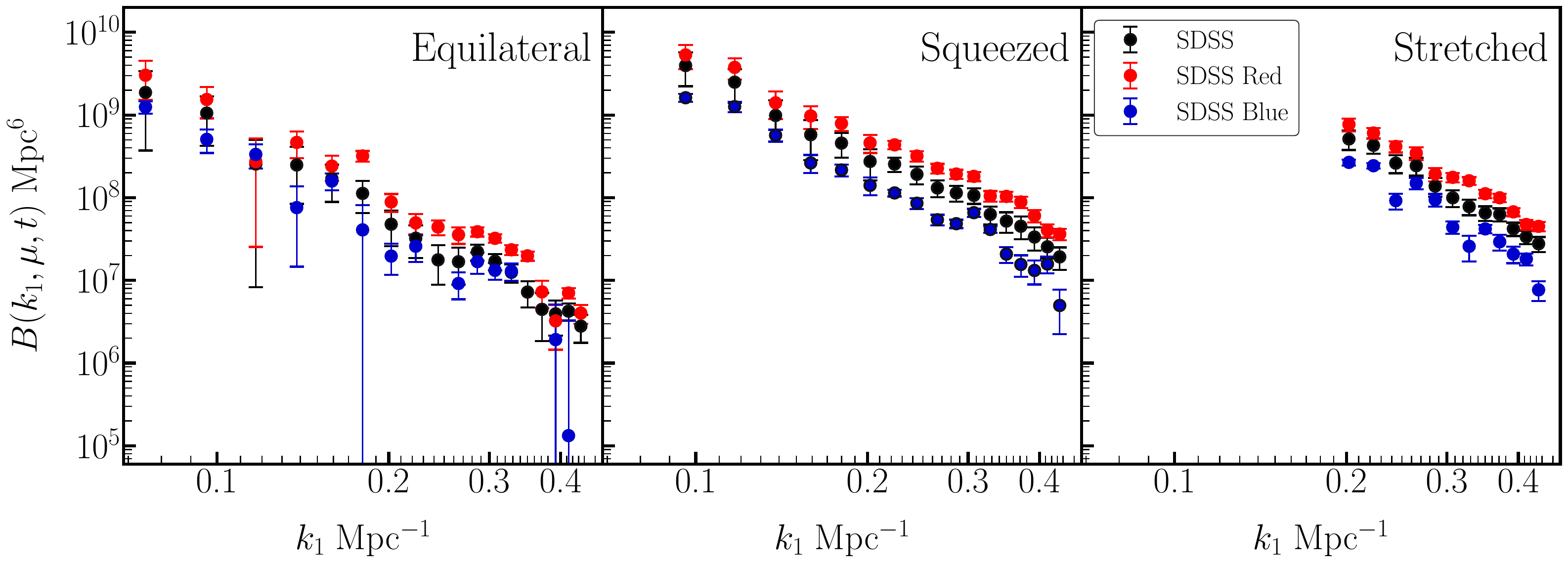}
    \caption{Similar to Figure~\ref{B_vs_k}, but for red and blue galaxies. The results considering the entire sample is also shown (black circles) for comparison.}
    \label{fig:B_vs_k_red_blue}
\end{figure}

We now investigate the clustering of red and blue galaxies. The right panel of Figure ~\ref{powsp} shows the power spectrum of red $\pkr$ and blue $\pkb$ galaxies. The general trend of $\pkr$ and $\pkb$ is similar to $\pks$ but with the difference in amplitude. $\pkr$ is seen to possess a higher amplitude than $\pkb$ throughout the entire $k$ range [$0.02 \, {\rm Mpc}^{-1}$, $0.7 \, {\rm Mpc}^{-1}$]. The red galaxies are found to be more strongly clustered and more biased than blue galaxies. Comparing with the left panel, we see that the blue galaxies have a smaller bias that is close to $b_1\approx 1$. Previous studies have found the same conclusion \citep{Zehavi2005,Zehavi2011,Skibba2013}. Figure~\ref{fig:B_vs_k_red_blue} shows the bispectrum of red $\bkr$ and blue $\bkb$ galaxies for equilateral, squeezed and stretched triangles. In this study, the errors are estimated using the jack-knife method. We prepare $10$ jack-knife samples, separately for red and blue galaxy data, by randomly picking $80\%$ galaxies from the original sample in each case. The $\bkr$ is larger in magnitude than $\bks$ and $\bkb$ in the same order for most of the cases. However, for a few cases in the equilateral triangles, the difference is not statistically significant as error bars of $\bkb$ are large. The reason can be the small sample size of blue galaxies that is incapable of capturing the large-scale non-Guassian features. The estimates and difference between the amplitudes of $\bkr$ and $\bkb$ at the larger $k_1$ range [$0.2 \,{\rm Mpc}^{-1}$,\, $0.4 \,{\rm Mpc}^{-1}$] are more robust for all triangle configurations. The bispectrum captures the non-linear mode coupling of underlying fluctuations. Red galaxies are typically older, having formed earlier in the cosmic history. Their larger bispectra suggests that their clustering has evolved through a series of complex and non-linear interactions within their environments over time. Blue galaxies, which are generally younger and more uniformly distributed, exhibit smaller bispectra. This indicates that their clustering is more linear and less influenced by complex gravitational interactions. Also, the blue galaxies are better tracers of the underlying dark matter field and, hence, are less biased.

\begin{figure}[htbp!]
    \begin{subfigure}{0.5\textwidth}
        \centering
        \includegraphics[width=\linewidth]{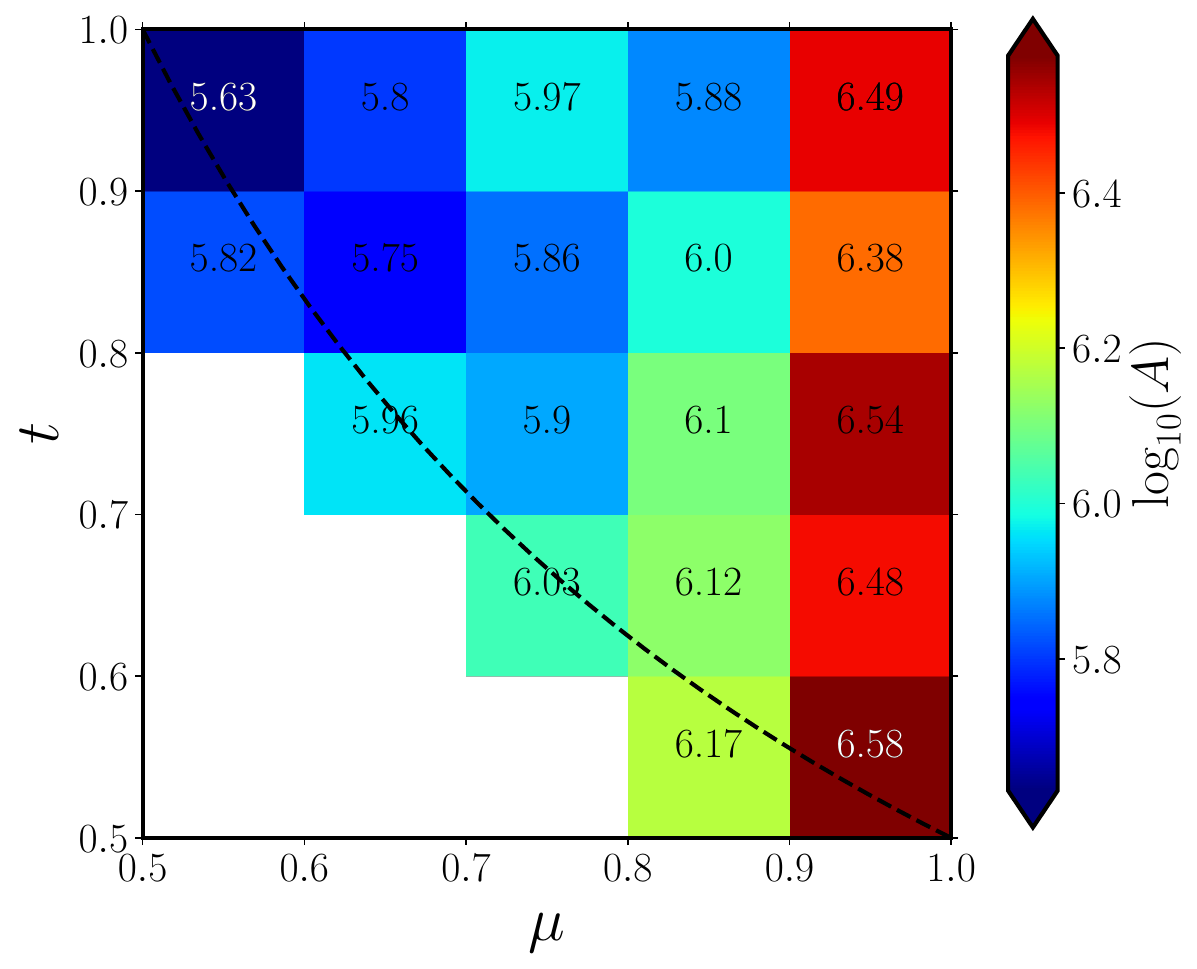}
    \end{subfigure}%
    \hspace{0.04\textwidth}
    \begin{subfigure}{0.5\textwidth}
        \centering
        \includegraphics[width=\linewidth]{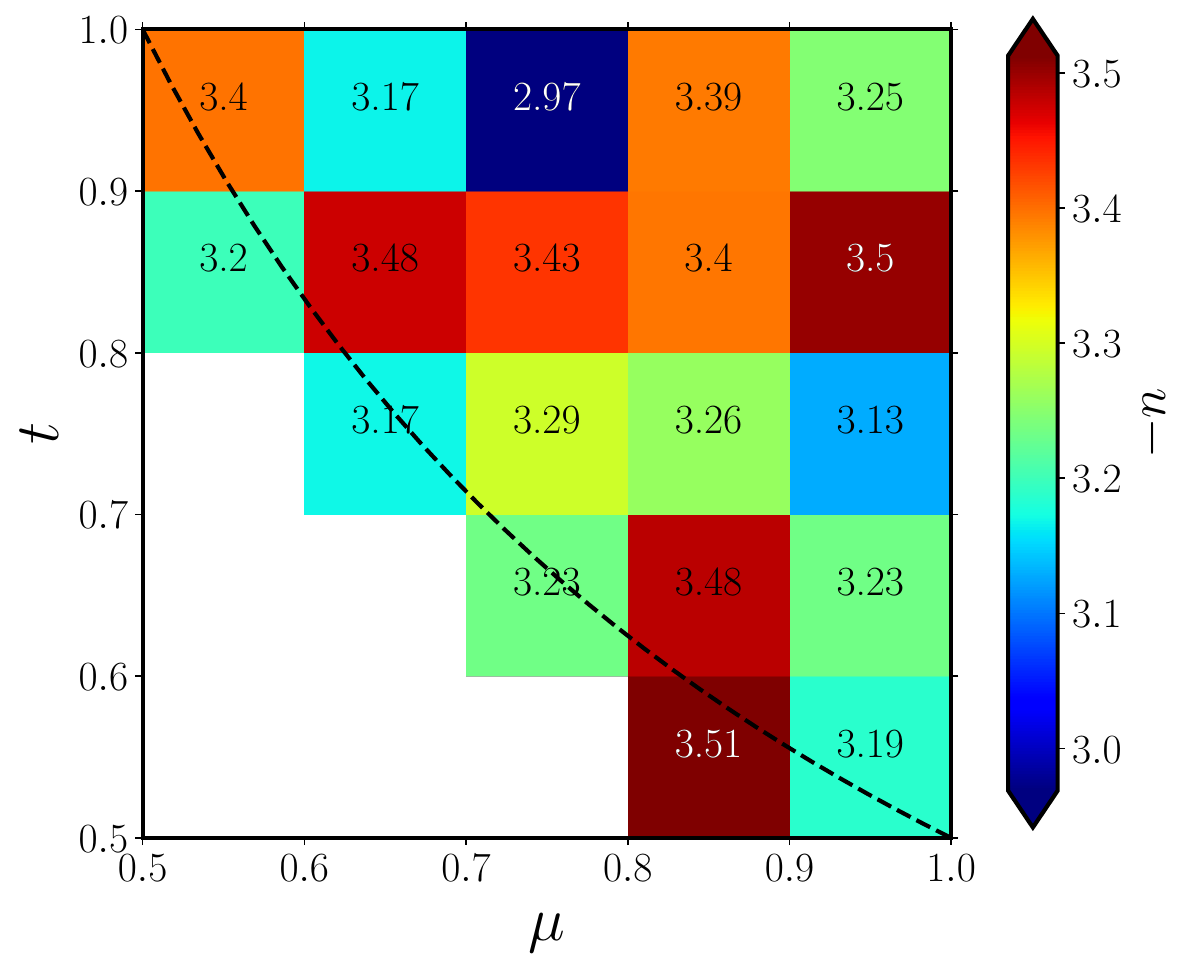}
    \end{subfigure}
\caption{Similar to Figure~\ref{fig:best_A_and_n}, but for red galaxies.}
\label{fig:best_A_and_n_for_red}
\end{figure}

\begin{figure}[htbp!]
    \begin{subfigure}{0.5\textwidth}
        \centering
        \includegraphics[width=\linewidth]{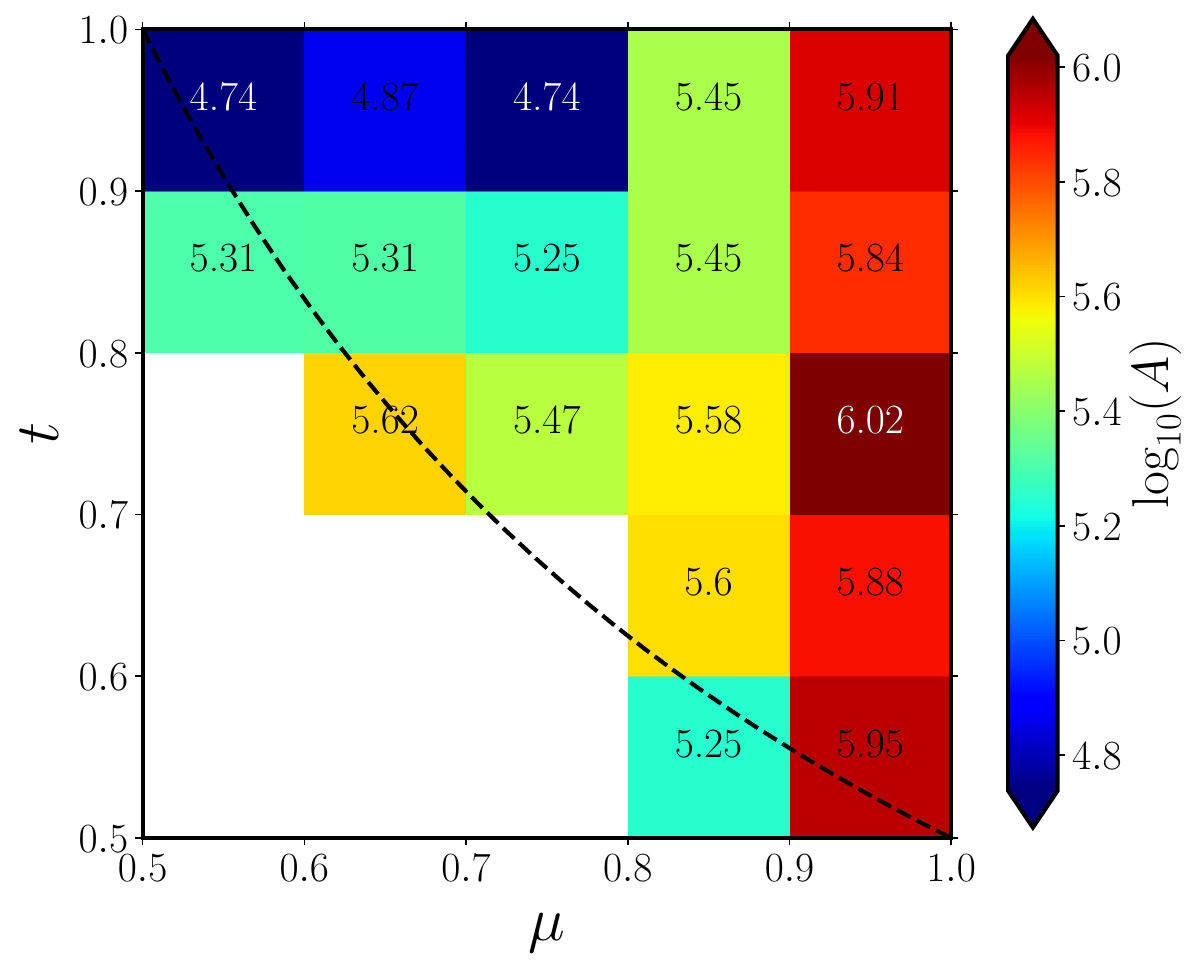}
    \end{subfigure}%
    \hspace{0.04\textwidth}
    \begin{subfigure}{0.5\textwidth}
        \centering
        \includegraphics[width=\linewidth]{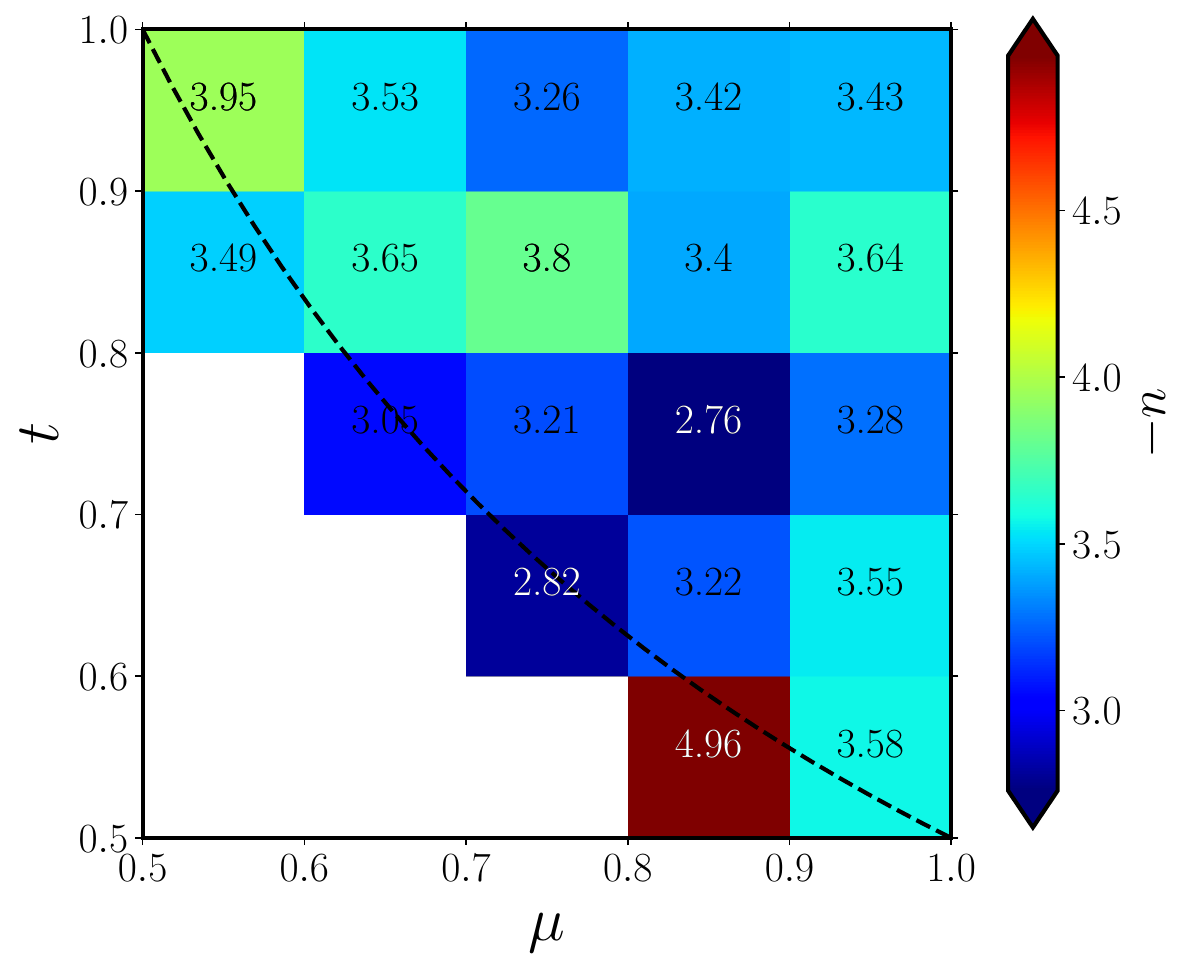}
    \end{subfigure}
\caption{Similar to Figure~\ref{fig:best_A_and_n}, but for blue galaxies.}
\label{fig:best_A_and_n_for_blue}
\end{figure}

Further, we fit the $\bkr$ and $\bkb$ with the power-law (Equation \ref{eq:fit}) to show size dependence for triangles of all possible shapes. The corresponding best-fit values of parameter $A$ and $n$ are shown in  Figures~\ref{fig:best_A_and_n_for_red} and \ref{fig:best_A_and_n_for_blue}  for the red and the blue galaxies, respectively. The variation of $A(\mu,t)$ is similar to the case of full SDSS data for both samples. The value of $A$ reaches its peak for linear triangles, particularly $\log_{10} A = 6.58$ at the stretched limit for red and $\log_{10} A = 6.02$ at bin $\mu = 0.95$, $t = 0.75$ for blue samples. As $\mu$ decreases with $t = 1$ fixed, $A$ declines sharply, with $\log_{10} A = 5.63$ and $4.74$ for red and blue galaxies at equilateral limit. Altogether, $A$ is relatively large for linear triangles ($\mu = 1$) and drops rapidly as we move from obtuse to acute triangles. We see that the amplitude $A$ is comparatively larger in the case of red galaxies corresponding to all triangle configurations. The variation of $n(\mu,t)$ in shape space is slightly different for the red galaxies than the blue galaxies. For red galaxies, the $\mid n \mid$ is minimum $(\mid n \mid = 2.97)$ for the bin $\mu = 0.75$, $t = 0.95$ and maximum $(\mid n \mid = 3.51)$ for the triangles with $\mu = 0.85, t = 0.55$. On the other hand, the minima $(\mid n \mid = 2.76)$ and maxima $(\mid n \mid = 4.96)$ for blue galaxies occur at $\mu = 0.85$, $t = 0.75$  and $\mu = 0.85$, $t = 0.55$,  respectively. We do not see any generic trend in the shape variation of slope $n(\mu ,t)$. For many triangle shapes, the slope is steeper for the blue galaxy bispectrum, particularly near linear and equilateral limit triangles. On the contrary, $\mid n\mid$ is larger for many bins at $0.7\leq\mu\leq0.9$ for red galaxy bispectrum.

\section{Summary and Discussion} 
\label{summary}

The statistical studies of the galaxy distribution, such as power spectrum, bispectrum etc., have been used for constraining the cosmology so far. For being a biased tracer of the underlying LSS, it can be potentially used to quantify the statistical nature of the underlying matter density field. Our aim here is to quantify the non-Gaussianity present in the LSS using the galaxy bispectrum. In this paper, we have estimated the monopole moment of the galaxy power spectrum and bispectrum using a $[296.75\, \rm Mpc]^3$ data cube with a mean galaxy number density of  $0.63 \times 10^{-3} \, {\rm Mpc}^{-3}$ and a median redshift of $0.102$. We use here the main galaxy sample \citep{2002AJStrauss} from SDSS \citep{2000AJYork}. Our analysis considers $\sim 1.37 \times 10^8$ triangles, for which we have analysed the binned bispectrum $B(k_1,\mu,t)$ to quantify the size $(k_1)$ and shape $(\mu,t)$ dependence.  The cubic volume and uniform selection criteria of galaxy data allow us to avoid several complications that would arise if we had a complex survey geometry and a spatially varying selection function (e.g. \cite{2021Philcox_cubic,2023Philcox_scalar,2023Philcox_tensor}), however, this restricts the $k$ range accessible for the analysis. The range of $k_1$ over which it is possible to estimate $\bks$ varies with $(\mu,t)$, and it spans the largest range $(0.075 - 0.434) \,{\rm Mpc}^{-1}$  for equilateral triangles.  We find that for every fixed shape $(\mu,t)$, the $k_1$ dependence of $\bks$ can be modelled as a power law $A \, (k_1/1 {\rm Mpc}^{-1})^n$ (equation~\ref{eq:fit}). The amplitude $A$ is minimum for equilateral triangles, and it increases steeply as the triangle shape is deformed to right-angled and then obtuse triangles. The amplitude is maximum for linear triangles where the two largest sides are aligned along nearly the same axis. The power law index $n$ is found to have negative values. The value of $|n|$ is maximum $(3.8)$ near the acute triangles, and it is minimum $(3.12)$ for S-isosceles triangles. Overall, $|n|$ decreases as we go from equilateral to linear triangles. 

\begin{figure*}[htb!]
    \includegraphics[width = 1\textwidth]{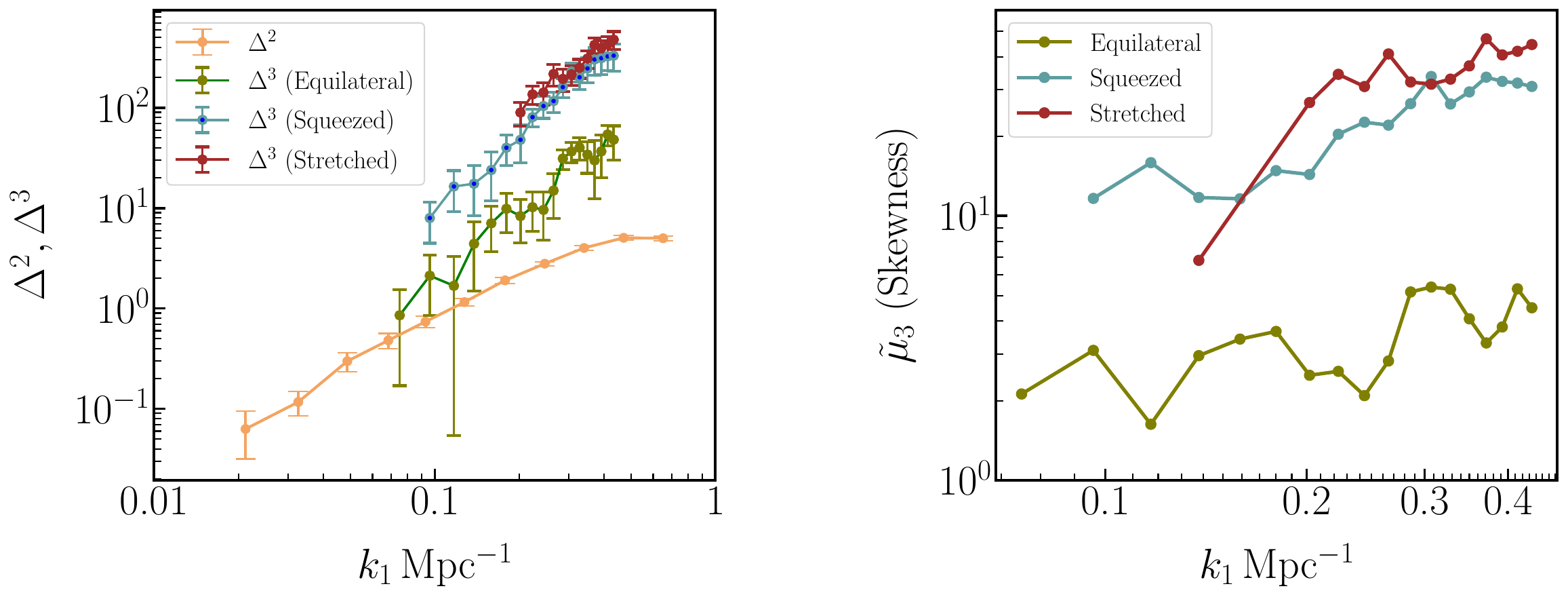}
    \caption{Dependence of dimensionless power spectrum, bispectrum and skewness on $k_1$. The left panel shows the dimensionless power spectrum $\Delta^2(k_1)=k_1^3 P(k_1)/(2\pi^2)$ and bispectrum $\Delta^3(k_1,\mu,t)=k_1^6 B(k_1,\mu,t)/(2\pi^2)^2$ estimated from SDSS galaxy data. The right panel shows the skewness parameter $\tilde{\mu}_3=\Delta^3/(\Delta^2)^{1.5}$ corresponding to equilateral, squeezed and stretched triangles.}
    \label{dl_psbs}
\end{figure*}

Considering the matter power spectrum, we know that it grows as
$P(k)\propto k^{1}$ at  $k<k_H$ where $k_H\,(\approx0.01\,\mpci)$ is the wave vector correspond to the horizon scale at the matter-radiation equality. As we approach $k \sim k_H$, the slope of the power spectrum gradually reaches zero. At $k>k_H$, the power spectrum decays with $k$ as $P \propto k^{n_p}$, where $n_p \approx -3$ at large $k$ typically for the linear power spectrum. For the non-linear power spectrum, $n_p > -3$. 
Analysing $[P(k)]_{\rm SDSS}$ (Figure~\ref{powsp}), we found that the slope of the power spectrum is $n_p\approx-2$ in the $k$ range $(0.03-0.47)\mpci$. Now, it is known that the bispectrum (at tree-order) scales as $B(k) \propto P(k)^2$. In order to relate the slope $n$ of the bispectrum (right panel of Figure~\ref{fig:best_A_and_n}) with $n_p$, we first focus on the equilateral triangles, $(\mu,t)\approx (0.5,1)$, where all the $k$-vectors have similar magnitude ($k_1 \approx k_2 \approx k_3$). Therefore, $B(k) \propto P(k)^2$ for the equilateral triangles in our analysis implies that $n \approx 2n_p \approx -4$, which is consistent with the best-fit value $-3.64$ that we have obtained. However, as we deviate from the equilateral configuration, the magnitudes of the three $k$ vectors differ, and we observe deviations from $n \approx -4$ for other triangles.

The bispectrum is the lowest-order statistic that is sensitive to the non-Gaussianity in the LSS. Although the linear primordial fluctuations are believed to be well described by a Gaussian random field, we expect the  LSS to be non-Gaussian at late times $(z \sim 0.1)$ due to the non-linear coupling of the different modes of the density fluctuations. The left panel of Figure~\ref{dl_psbs} shows the dimensionless quantity $\Delta^2(k_1)=k_1^3 P(k_1)/(2\pi^2)$ which provides an estimate of the mean squared fluctuations of the observed density contrast in Fourier space. We see that the fluctuations are significantly non-linear $(\Delta^2(k_1) > 1)$ at the length-scales $k_1 > 0.075 \, {\rm Mpc}^{-1}$ where we have measured the bispectrum, and  $\Delta^2(k_1)$ increases  from $0.74$ to $4.01$  across the $k_1$ range $(0.075  -  0.434 )\, {\rm Mpc}^{-1}$. This panel also shows the dimensionless quantity $\Delta^3(k_1,\mu,t)=k_1^6 B(k_1,\mu,t)/(2\pi^2)^2$ for three different triangle shapes, we may interpret this as the mean cubed fluctuations of the observed density contrast in Fourier space. We see that $\Delta^3(k_1,\mu,t)$ exceeds $\Delta^2(k_1)$ throughout, and its values span the intervals $[ 0.86  , 48.06]$, $[ 7.96  , 330.19 ]$ and $[ 10.12  , 475.77 ]$ for the equilateral, squeezed and stretched triangles respectively. We see that the $\Delta^3(k_1,\mu,t)$ values are an order of magnitude larger for the linear (squeezed and stretched) triangles  in comparison to the equilateral triangles.  Further, $\Delta^3(k_1,\mu,t)$ also increases more steeply with $k_1$ for the linear triangles as compared to the equilateral triangles. 

The right panel of Figure~\ref{dl_psbs} shows  the skewness  $\tilde{\mu}_3=\Delta^3/(\Delta^2)^{1.5}$ which is a dimensionless measure of the deviations from a symmetric Gaussian distribution for the density contrast. A value $\tilde{\mu}_3 >1$  indicates a highly skewed distribution with a long positive tail. We see that in all cases we have a highly skewed distribution with $\tilde{\mu}_3 >1$. Considering equilateral triangles, $ \tilde{\mu}_3 $ has values in the range $\sim 2$ to $5$, and it does not exhibit any systematic $k_1$ dependence. In contrast, for the linear triangles, we see that  $ \tilde{\mu}_3 $ has much larger values in the range $6$ to $50$, and it increases with $k_1$. 

In conclusion of the present discussion we note that the observed galaxy distribution is non-linear and highly non-Gaussian at the intermediate scales probed in this analysis. Further, the degree of non-Gaussianity depends on the shape of the triangle, and it is minimum for equilateral triangles and maximum when the triangle is linear {\it i.e.} the  two largest sides are nearly aligned in the same direction. The bispectrum is the lowest order statistic which is sensitive to the shape of the structures in the Universe. The shape dependence of the bispectrum (and the non-Gaussianity) possibly arises from the sheets and filaments observed in the galaxy distribution, however we do not venture to make a definite quantitative statement regarding this here.  

Here we consider mock galaxy samples by using $\Lambda$CDM N-body simulations along with a simple Eulerian biasing scheme \citep{1998MNRASCole} where the galaxies reside in regions with the smoothed density exceeding a threshold.  We find that the bispectrum for the mock sample with bias $b_1=1.2$ is in good agreement with the SDSS bispectrum measured here. The biasing scheme has only two free parameters, namely the smoothing length-scale and the value of the density threshold. In future work, we plan to measure the higher multipole moments of the bispectrum, and combine these with simulations to address cosmological parameter estimation. 

Our analysis of red and blue type galaxy clustering supports the conclusions in the existing literature. The red galaxies exhibit a greater power spectrum and bispectrum for all triangle configurations than blue galaxies. The red galaxies are typically older and found in denser environments within high-mass halos. They are more common in galaxy clusters and have a stronger non-linear clustering signal compared to blue galaxies, which are more uniformly distributed. The blue galaxies better follows the underlying dark matter distribution with $b_1\approx 1$. We plan to extend this study with a larger galaxy sample in a future work.

\section*{Acknowledgements}
AN acknowledges the financial support from the Department of Science and Technology (DST), Government of India through an INSPIRE fellowship. SSG acknowledges the support of the Prime Minister's Research Fellowship (PMRF). AKS acknowledges support by the Israel Science Foundation (grant no. 255/18). DS acknowledges the support of the Canada 150 Chairs program, 
the Fonds de recherche du Québec Nature et Technologies (FRQNT) and Natural Sciences and Engineering Research Council of Canada (NSERC) joint NOVA grant, and the Trottier Space Institute Postdoctoral Fellowship program.
BP would like to acknowledge financial support from the SERB, DST, Government of India through the project CRG/2019/001110. BP would also like to acknowledge IUCAA, Pune, for providing support through the associateship programme. The authors thank the anonymous reviewer for the valuable comments, which helped to improve the quality of this work.

\bibliographystyle{elsarticle-harv} 
\bibliography{ref}

\appendix

\section{Appendix}
\label{sec:app}
\begin{figure*}
    \centering
    \hspace{-1.6cm} 
    \includegraphics[width=16cm]{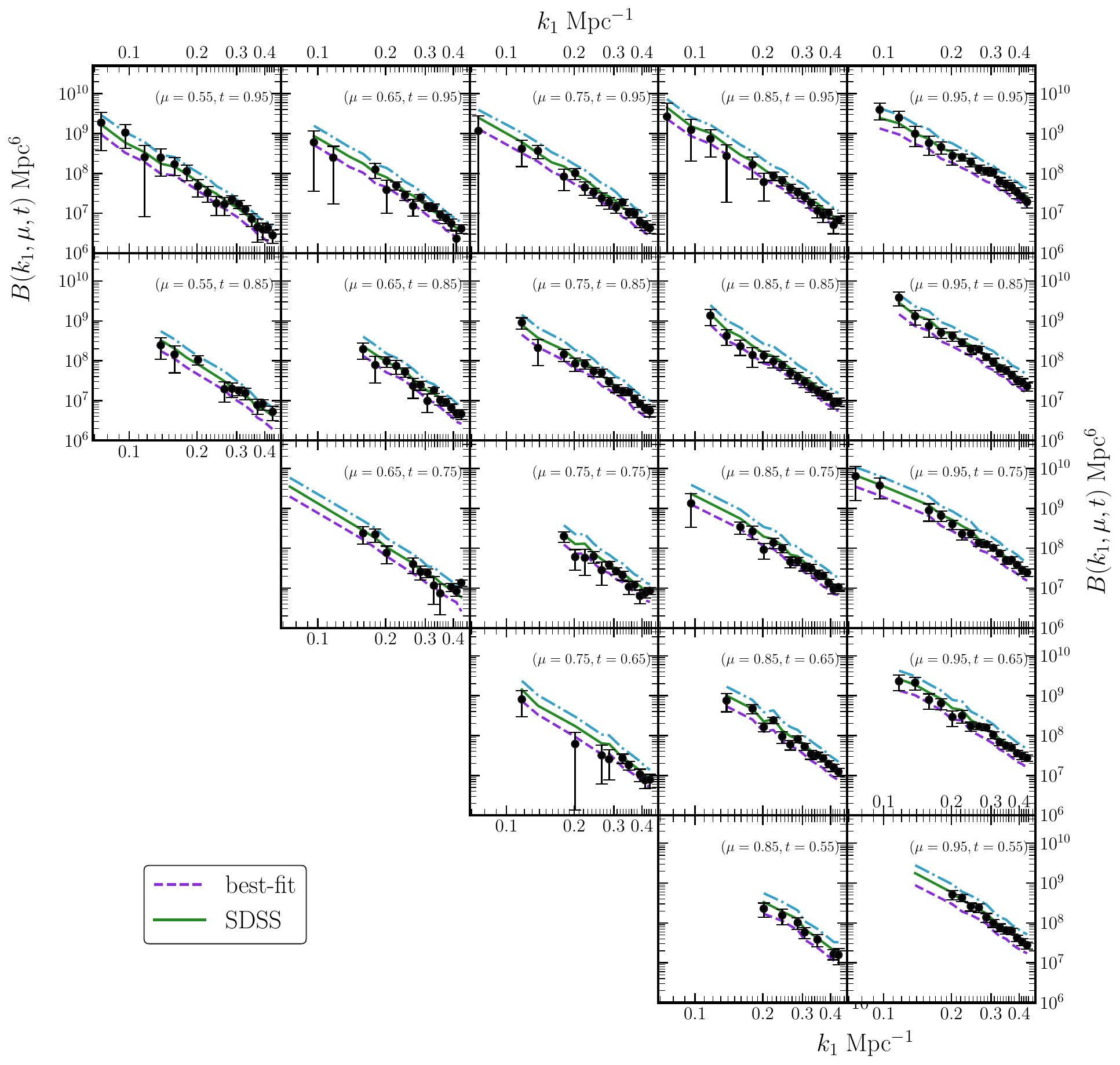}

    \caption{Bispectrum $B(k_1,\mu,t)$  as a function of $k_1$. Here each panel corresponds to a different $(\mu,t)$ bin.    The values of $\mu$ and $t$ increase from $0.55$ to $0.95$ along the horizontal and vertical directions respectively.  In addition to the SDSS, results are also shown for mock galaxy samples with $b_1=1,~1.2$ and $1.4$ respectively. The error bars correspond to the $1 \sigma$ errors for $b_1=1.2$ which provides a good match for the SDSS data.}
    \label{fig:SDSS_bias_all_shapes}
\end{figure*}

\begin{figure*}
    \centering
    \hspace{-1.63cm}
    \includegraphics[width=17cm]{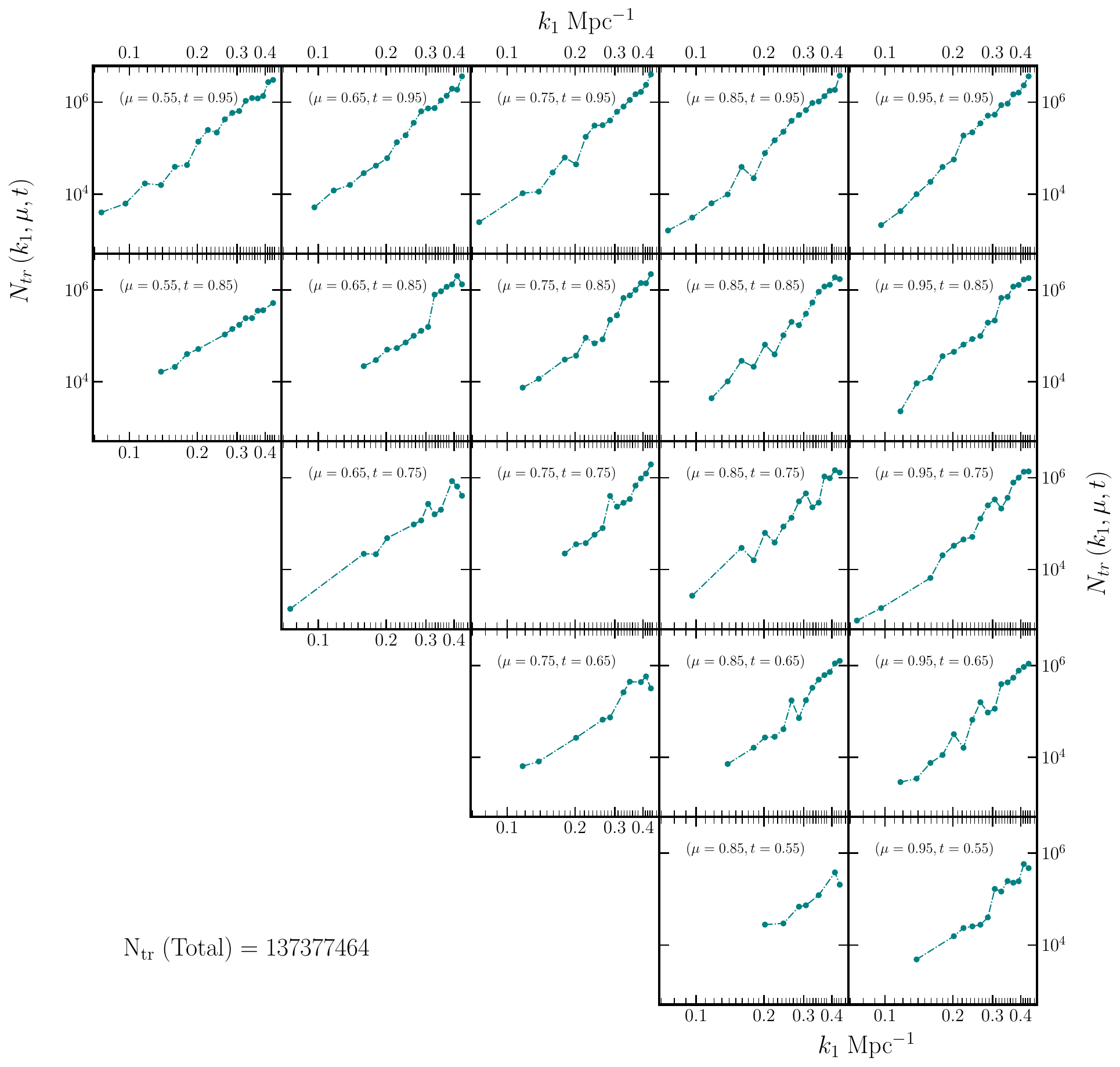}
    \caption{Similar to Figure~\ref{fig:SDSS_bias_all_shapes}, except that the panels now show $N_{tr}(k_1,\mu,t)$ the number of triangles in each $(k_1,\mu,t)$  bin.}
    \label{fig:SDSS_ntri_all_shapes}
\end{figure*}

Figure \ref{fig:SDSS_bias_all_shapes} shows the size ($k_1$) dependence of the SDSS galaxy bispectrum $\bks$ for triangles of different shapes.  As discussed in Section~\ref{method}, each value of $(\mu,t)$ corresponds to a different triangle shape. Here we have divided the $(\mu,t)$ plane into 
$5\times5$  linear bins, of these nineteen bins cover the allowed region of the $(\mu,t)$ plane. Each panel of Figure \ref{fig:SDSS_bias_all_shapes} shows the $k_1$ dependence of $\bks$ for a fixed  $(\mu,t)$ bin. Each panel also shows $[B(k_1,\mu,t)]_{b_1}$ estimated from the mock galaxy samples for the three bias values $b_1=1,~1.2$ and $1.4$ respectively.  The error bars plotted on the SDSS results correspond to $1-\sigma$ errors for $b_1=1.2$, which provides a good match for the SDSS results.  Figure~\ref{fig:SDSS_ntri_all_shapes}  shows the corresponding number of triangles in each $(k_1,\mu,t)$ bin.  We have a total $N_{tr}({\rm Total})=137,377,464$ triangles which are included in our analysis. 

\begin{figure*}
    \centering
    \hspace{-1.63cm}
    \includegraphics[width=17cm]{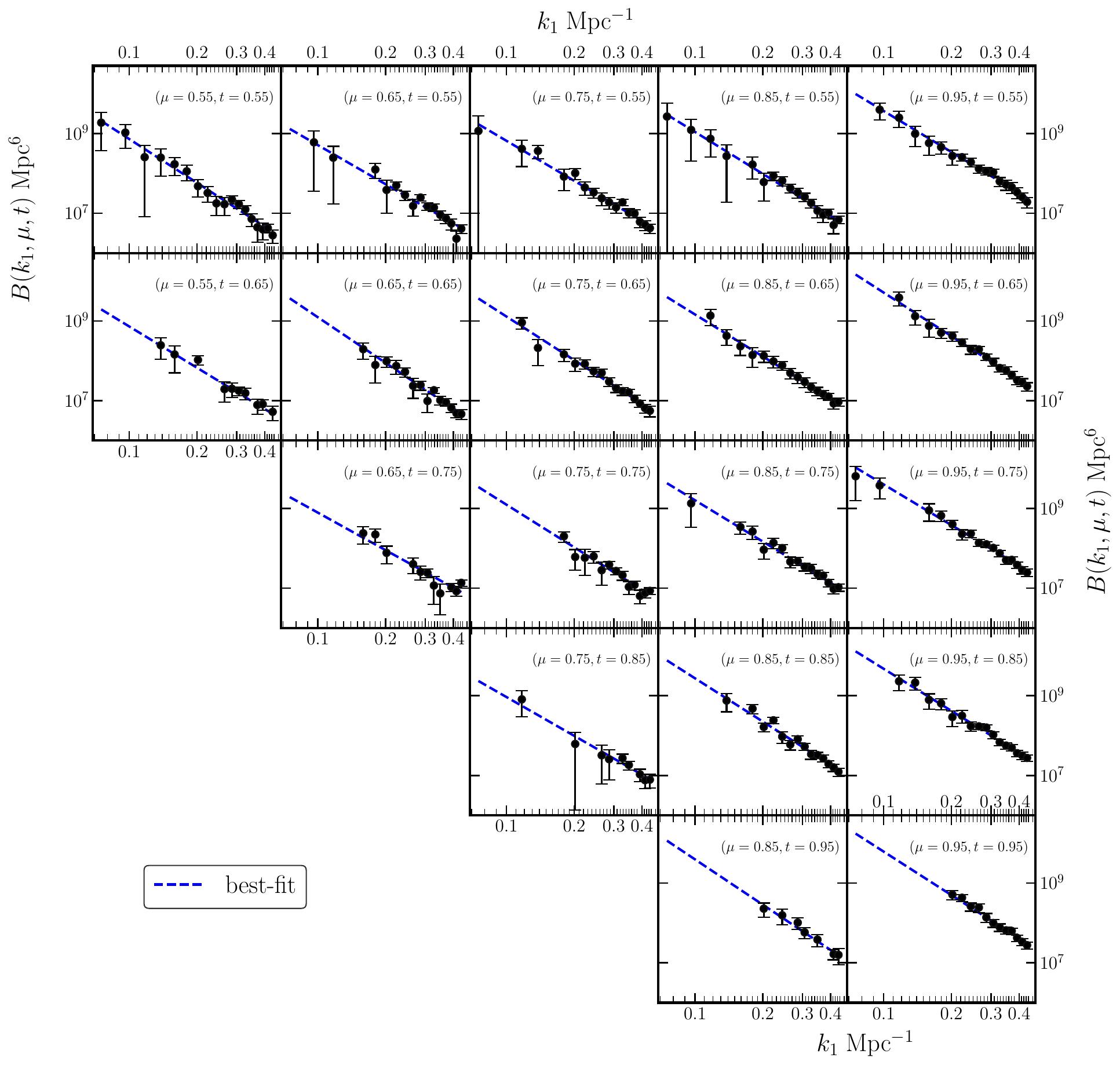}

    \caption{Similar to Figure~\ref{fig:SDSS_bias_all_shapes}, except that the  panels now show  $\bks$ along with the best-fit power law  $B(k_1,\mu,t)= A\,\big(k_1/1\mpci\big)^{n}$ where  the  best-fit parameter values for  $A$ and $n$ vary with $(\mu,t)$. }
    \label{fig:SDSS_bias_all_shapes_fit}
\end{figure*}

We find that in each $(\mu,t)$ bin (Figure \ref{fig:SDSS_bias_all_shapes}), the $k_1$ dependence of $\bks$ is well fitted by a power law   $B(k_1,\mu,t)= A\,\big(k_1/1\mpci\big)^{n}$. The best-fit values of the parameters $A$ and $n$ vary with $(\mu,t)$. The different panels of Figure~\ref{fig:SDSS_bias_all_shapes_fit} show $\bks$ along with the corresponding best-fit power law. The best-fit parameter values and the corresponding reduced chi-square ($\bar{\chi^2}$) values are tabulated in Table~\ref{tab:SDSS_fit_for_diff_mu_t}. 


\renewcommand{\arraystretch}{1.5} 
\definecolor{darkbrown}{rgb}{0.4, 0.26, 0.13}

\newcommand\mcl[1]{\multicolumn{2}{|l|}{#1}}
\begin{table}[htb!]

    \centering
    \begin{tabular}{|c|c|c|c|c|}
    \hline
    
  $\mu$  & $t$  &\multirow{1}{*}{$A~({\rm Mpc}^6)$} & \multirow{1}{*}{$n$} & \multirow{1}{*}{$\bar{\chi^2}$ } \\
     \hline
     $0.85$ & $ 0.55$ & $(6.65 \pm 3.86)\times 10^5$ & $(-3.77 \pm 0.49)$ & $0.29$ \\
     \hline
     \textcolor{darkbrown}{$ 0.95$} & \textcolor{darkbrown}{$0.55$}  & \textcolor{darkbrown}{$(1.47 \pm 0.41)\times 10^6$} & \textcolor{darkbrown}{$(-3.62 \pm 0.24)$} & \textcolor{darkbrown}{$0.61$} \\
     \hline
      $ 0.75$&$ 0.65$& $(5.44 \pm 2.85)\times 10^5$ & $(-3.23 \pm 0.47)$ & $0.33$ \\
     \hline
     $ 0.85$&$ 0.65$ & $(6.85 \pm 1.81)\times 10^5$ & $(-3.60 \pm 0.22)$ & $0.96$ \\
     \hline
     $ 0.95$&$ 0.65$ & $(1.48 \pm 0.33)\times 10^6$ & $(-3.50 \pm 0.18)$ & $0.45$ \\
     \hline
     $ 0.65$&$ 0.75$ & $(5.83 \pm 2.34)\times 10^5$ & $(-3.12 \pm 0.35)$ & $1.12$ \\
     \hline
     $ 0.75$&$ 0.75$ & $(3.48 \pm 1.27)\times 10^5$ & $(-3.55 \pm 0.32)$ & $0.86$ \\
     \hline
     $ 0.85$&$0.75$ & $(5.81 \pm 1.55)\times 10^5$ & $(-3.44 \pm 0.22 )$ & $0.56$ \\
     \hline
     $ 0.95$&$ 0.75$ & $(1.62 \pm 0.34)\times 10^6$ & $(-3.39 \pm 0.17)$ & $0.34$ \\
     \hline
     $ 0.55$&$ 0.85$ & $(2.64 \pm 1.32)\times 10^5$ & $(-3.44 \pm 0.39)$ & $0.94$ \\
     \hline
     $ 0.65$&$ 0.85$ & $(1.96 \pm 0.64)\times 10^5$ & $(-3.80 \pm 0.28)$ & $0.70$ \\
     \hline
     $ 0.75$&$ 0.85$ & $(3.04 \pm 0.76)\times 10^5$ & $(-3.63 \pm 0.20)$ & $0.46$ \\
     \hline
     $ 0.85$&$ 0.85$ & $(4.63 \pm 1.23)\times 10^5$ & $(-3.50 \pm 0.21)$ & $0.21$ \\
     \hline
     $ 0.95$&$ 0.85$ & $(1.22 \pm 0.29)\times 10^6$ & $(-3.63 \pm 0.18)$ & $0.28$ \\
     \hline
     \textcolor{olive}{$ 0.55$}&\textcolor{olive}{$ 0.95$} & \textcolor{olive}{$(1.65 \pm 0.51)\times 10^5$ }& \textcolor{olive}{$(-3.64 \pm 0.24)$} & \textcolor{olive}{$0.53$} \\
     \hline
     $ 0.65$&$ 0.95$ & $(2.84 \pm 0.90)\times 10^5$ & $(-3.26 \pm 0.26)$ & $1.19$ \\
     \hline
     $ 0.75$&$ 0.95$ & $(3.06 \pm 0.91)\times 10^5$ & $(-3.33 \pm 0.24)$ & $0.85$ \\
     \hline
     $0.85$&$ 0.95$ & $(3.56 \pm 1.04)\times 10^5$ & $(-3.49 \pm 0.24)$ & $0.61$ \\
     \hline
    \textcolor{teal}{ $ 0.95$}&\textcolor{teal}{$ 0.95$} & \textcolor{teal}{$(1.40 \pm 0.36)\times 10^6$} & \textcolor{teal}{$(-3.42 \pm 0.19)$} & \textcolor{teal}{$0.27$} \\
     \hline
     
    \end{tabular}
    \caption{Using a power law $B(k_1,\mu,t)= A\,\big(k_1/1\mpci\big)^{n}$  to fit $\bks$, this tabulates the best-fit parameter values for $A$ and $n$, and the corresponding reduced chi-square  $\bar{\chi^2}$  for each $(\mu,t)$ bin considered in Figure~\ref{fig:SDSS_bias_all_shapes_fit}. Results for stretched, equilateral, and squeezed triangles are highlighted with red, blue, and green, respectively.}
    \label{tab:SDSS_fit_for_diff_mu_t}

\end{table}

\newpage

\section{Appendix}
\begin{figure}[htbp!]
    \centering
    \begin{subfigure}[b]{\textwidth}
    \centering
    \includegraphics[width=0.79\linewidth]{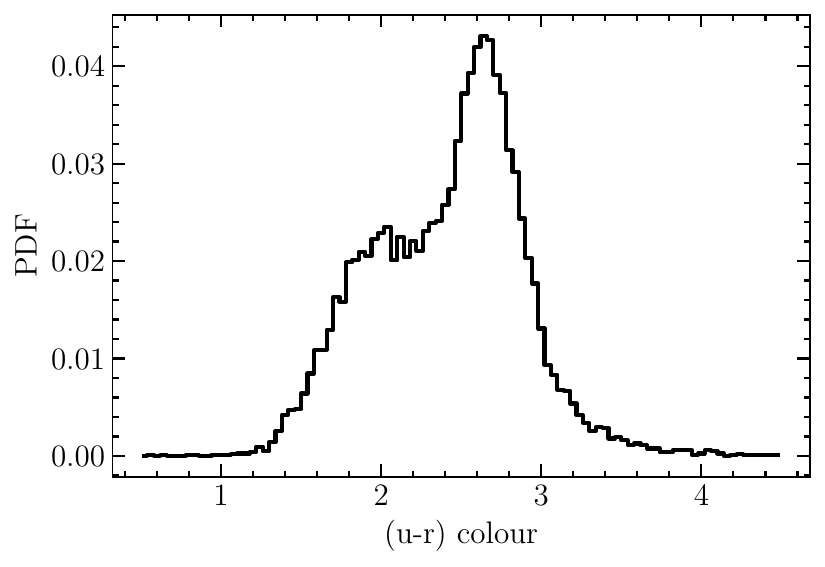}
    \caption{}
    \label{fig:color_pdf}
    \end{subfigure}%
    \vspace{1cm}
    \begin{subfigure}[b]{\textwidth}
    \centering
    \includegraphics[width=0.75\linewidth]{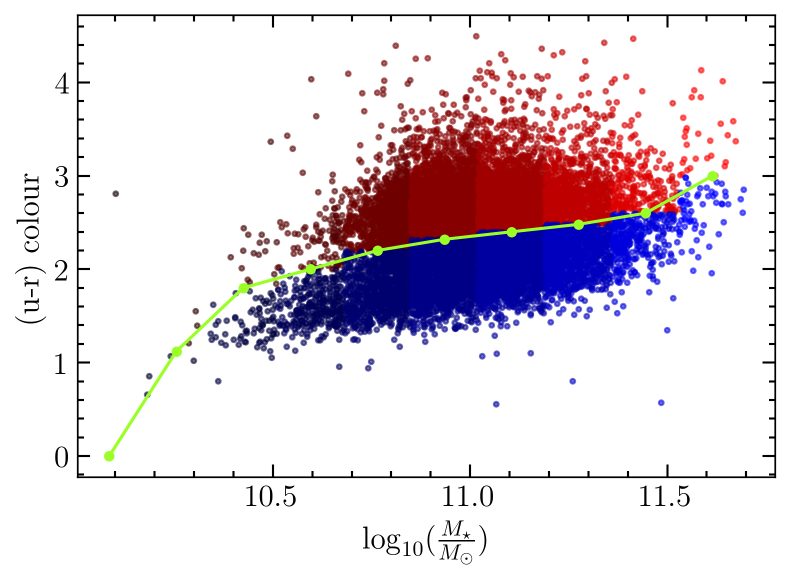}
    \caption{}
    \label{fig:red_blue_classification}
    \end{subfigure}   
    \caption{(a) Distribution of (u-r) colour of the galaxies in the SDSS data cube. (b) Classification of red and blue galaxies in the (u-r) color-$\log_{10}(\frac{M_{\star}}{M_{\odot}})$ plane.}
\end{figure}







\end{document}